\documentstyle[12pt,epsf]{article}
\begin{document}
\centerline{\large\bf CP violation and matter effect in long-baseline }
\vskip 0.4truecm
\centerline{\large\bf neutrino oscillation in the four-neutrino model}
\baselineskip=8truemm
\vskip 0.8truecm
\centerline{Toshihiko Hattori$,^{a),}$\footnote{e-mail:
hattori@ias.tokushima-u.ac.jp} \ Tsutom Hasuike$,^{b),}$\footnote{e-mail:
hasuike@anan-nct.ac.jp} \ and \ Seiichi Wakaizumi$ ^{c),}$\footnote{e-mail:
wakaizum@medsci.tokushima-u.ac.jp}}
\vskip 0.5truecm
\centerline{\it $ ^{a)}$Institute of Theoretical Physics, University of
Tokushima,
Tokushima 770-8502, Japan}
\centerline{\it $ ^{b)}$Department of Physics, Anan College of Technology,
Anan 774-0017, Japan}
\centerline{\it $ ^{c)}$School of Medical Sciences, University of Tokushima,
Tokushima 770-8509, Japan}
\vskip 0.8truecm
\centerline{\bf Abstract}
\vskip 0.5truecm

We investigate the CP violation effect and the matter effect in the
long-baseline neutrino oscillations in the four-neutrino model with mass
scheme of the two pairs of two close masses separated by a gap of the order
of 1 eV by using the constraints on the mixing matrix derived from the solar
neutrino deficit, atmospheric neutrino anomaly, LSND experiments, and the
other terrestrial neutrino oscillation experiments. We also use the results
of the combined analyses by Gonzalez-Garcia, Maltoni, and Pe\~na-Garay
of the solutions to the solar and atmospheric neutrino problems with the
recent SNO solar neutrino data. For the solution of close-to-active solar 
neutrino oscillations 
plus close-to-sterile atmospheric neutrino oscillations, the pure CP
violation part of the oscillation probability difference between the
CP-conjugate channels could attain as large as $0.10-0.25$ in the neutrino
energy range of $E = 6-15$ GeV at the baseline $L = 730$ km for
$\nu_\mu\to\nu_\tau$ oscillation and the matter effect is at the $8-15\%$
level of the pure CP violation effect, while for the solution of
near-pure-sterile solar neutrino oscillations plus near-pure-active atmospheric neutrino
oscillations, the pure CP violation effect in $\Delta P(\nu_\mu\to\nu_\tau)$
is very small $(\sim 0.01)$ and is comparable to the matter effect.
For $\nu_\mu\to\nu_e$ oscillation, the pure CP violation effect is
independent
of the active-sterile admixture and is at most $0.05$ in $E = 1.5-3$ GeV at
$L = 290$ km and the matter effect is at the $15-30\%$ level.
\newpage

\centerline{\large\bf I  Introduction}
\vskip 0.2truecm

CP violation is not yet observed in the leptonic sector. Since it is found
in the hadronic sector such as in $K$ \cite{Christenson} and $B$ meson
\cite{Aubert} decays, the observation of CP violation in the neutrino
oscillation will bring an important clue to understand the origin of CP
violation.

The solar neutrino deficit \cite{Lande} and the atmospheric neutrino anomaly
\cite{Fukuda98} have been interpretted as an evidence of the neutrino
oscillation. The relevant mass-squared differences of the neutrinos are
derived to be $\Delta m^2_{{\rm atm}} = (1.5-5)\times 10^{-3}$ ${\rm eV}^2$
for the atmospheric neutrino anomaly \cite{FukudaPRL98} and to be in the
range
$\Delta m^2_{{\rm solar}} = (10^{-11} - 10^{-4})$ ${\rm eV}^2$,
corresponding
to the four solutions to the solar neutrino deficit \cite{Bahcall}.
Moreover,
the Liquid Scintillation Neutrino Detector (LSND) measurements \cite{LSND}
has given a possible evidence of $\nu_\mu\to\nu_e$ and $\bar{\nu_\mu}
\to\bar{\nu_e}$ oscillations with $\Delta m^2_{{\rm LSND}} = (0.2-1)$
${\rm eV}^2$ in the short-baseline experiments.

The recent measurement of the solar neutrino flux by the use of $\nu_e$
charged current process on deuteron disintegration by the Sudbury Neutrino
Observatory (SNO) \cite{SNO} seems to indicate that the large mixing angle
(LMA) solution and the low mass (LOW) solution in the MSW mechanism survive
among the neutrino oscillation solutions for the solar neutrino problem in
the
three-neutrino mixing scheme \cite{Fogli}.

As for the sterile neutrino [10-19], the oscillation into sterile neutrinos
is claimed to be disfavored by the Super-Kamiokande Collaboration for both
the
solar neutrino \cite{Fukuda01} and the atmospheric neutrino \cite{Fukuda00}
transitions in the two-neutrino analyses. However, the recent four-neutrino
analyses by Barger, Marfatia, and Whisnant \cite{Marfatia} and by
Gonzalez-Garcia, Maltoni, and Pe\~na-Garay \cite{Gonzalez-Garcia} including
the SNO measurement show that the oscillation into the active-sterile
admixture is allowed for both the solar neutrino and the atmospheric
neutrino.

CP violation in the long-baseline neutrino oscillations has been
investigated
in the three-neutrino mixing scheme [24-29], including the earth matter
effect
\cite{Wolfenstein}. The size of the CP violation effects turns out to be
of $({\rm a\,\, few}-10)\%$
level up to the neutrino energy $E \sim 1$ ${\rm GeV}$ at a baseline $L =
250-730$ km for the mass-squared differences $\Delta m^2_{21} \equiv
\Delta m^2_{\rm solar} \simeq 3\times 10^{-5}{\rm eV}^2$, $\Delta
m^2_{31} \equiv \Delta m^2_{\rm atm} \simeq 3\times 10^{-3}{\rm eV}^2$ and
$|U_{e3}| \simeq 0.05$ which is related to the only undetermined angle at
present of the three mixing angles \cite{Sato}. The matter effect affects
the
pure CP violation effect, depending on the length of the baseline, although
there are cases in which the oscillation probabilities are approximately
independent of the presence of matter, called as "vacuum mimicking
phenomena"
\cite{Lipari}. The observability of the CP violation effects in
long-baseline
experiments is recently extensively studied for both the beams from the
neutrino factory \cite{Rujula} and the conventional superbeams \cite{Geer}.

On the other hand, the CP violation effect in the four-neutrino mixing
scheme
with one sterile neutrino is shown to be possibly highly sizable
\cite{Bilenky}
and is studied by us about its dependence on the mixing angles and
phases for various oscillations such as $\nu_e\to\nu_\mu $ and $\nu_\mu
\to\nu_\tau $ and is shown to reach to the magnitude as large as 0.3 in the
$\nu_\mu\to\nu_\tau $ oscillation in the long-baseline experiments
\cite{Hattori}. Since the oscillation pattern is governed by the LSND mass
scale in the short-baseline experiments in the four-neutrino mixing scheme,
the sensitivity to CP violation at the neutrino factory is studied in detail
at the baseline $L = 10-100$ km \cite{Barger01}\cite{Donini}.

In this paper we will investigate in the four-neutrino model how large the
CP
violation effects can be in the long-baseline experiments with $L = 290$ km
and 730 km for Gonzalez-Garcia et al's two solutions \cite{Gonzalez-Garcia},
that is, (A) close-to-active solar neutrino oscillations plus
close-to-sterile
atmospheric neutrino oscillations and (B) near-pure-sterile solar neutrino
oscillations plus near-pure-active atmospheric neutrino oscillations, and
will evaluate the matter effect to these CP violation effects in the
four-neutrino mixing scheme.

The paper is organized as follows. In Sec. II the method of calculating the
oscillation probability with matter effect, formulated
by Arafune, Koike, and Sato \cite{Arafune}, is applied to the four-neutrino
model. In Sec. III the constraints on the four-neutrino mixing matrix are
derived by using the results from the recent combined analyses of the solar
and atmospheric neutrino deficits in the four-neutrino scheme
\cite{Gonzalez-Garcia} and using the data from the LSND, Bugey, CHOOZ,
CHORUS,
and NOMAD experiments. In Sec. IV we study the behavior of CP violation
effect
with respect to the mixing angle that governs the active-sterile admixture
and, therefore, distinguishes the above-mentioned
two solutions (A) and (B), and we show our results on the pure CP violation
effects and the matter effect in the long-baseline experiments for
$\nu_\mu\to\nu_e$ and $\nu_\mu\to\nu_\tau$ oscillations with the baselines
of
$L = 290$ km and 730 km. It turns out that the CP violation effect in
$\nu_\mu\to\nu_\tau$ oscillation can be highly sizable $(\sim 0.2)$ for the
solution (A) and is very small $(\sim 0.02)$ for the solution (B). On the
other hand, the matter effect is small for $\nu_\mu\to\nu_e$ oscillation in
the neutrino energy range $E \le 2$ ${\rm GeV}$ at $L = 290$ km. For
$\nu_\mu\to\nu_\tau$ oscillation, the matter effect is negligibly small in
$E \le 12$ ${\rm GeV}$ at $L = 730$ km for the solution (A), while it is
comparable to pure CP violation effect for the solution (B). Sec. V is
devoted
to the conclusion.
\vskip 0.7truecm

\centerline{\large\bf II  Oscillation probability in the four-neutrino
model}
\vskip 0.2truecm

In order to consider the solar neutrino deficit, the atmospheric neutrino
anomaly and the LSND experiment, we take the
four-neutrino model with the three ordinary active neutrinos and one sterile
neutrino with three different scales of the neutrino mass-squared
difference,
$\Delta m^2_{{\rm solar}} = (10^{-6} - 10^{-4})$ ${\rm eV}^2,
\Delta m^2_{{\rm atm}} = (1.5-5)\times 10^{-3}$ ${\rm eV}^2$, and
$\Delta m^2_{{\rm LSND}} = (0.2-1)$ ${\rm eV}^2$.

Under the notion of the neutrino oscillation hypothesis \cite{Maki}
\cite{Pontecorvo}, the flavor eigenstates of neutrinos $\nu_\alpha\; (\alpha
=
e, \mu, \tau, s)$ are the mixtures of mass eigenstates in the vacuum
$\nu_i\;
(i = 1, 2, 3, 4)$ with masses $m_i$  as follows:
\begin{equation}
\nu_\alpha = \sum_{i=1}^4 U^{(0)}_{\alpha i}\, \nu_i ,    \label{shiki1}
\end{equation}

\noindent
where $\nu_e, \nu_{\mu}$ and $\nu_\tau$ are the ordinary neutrinos and
$\nu_s$ is the sterile neutrino, and $U^{(0)}$ is the unitary mixing
matrix.

In order to evaluate the matter effect, which gives a fake CP violation
effect, in the long-baseline neutrino oscillation experiments, we apply the
method formulated by Arafune, Koike, and Sato \cite{Arafune} to the
four-neutrino mixing scheme.

The evolution equation for the flavor eigenstate vector in matter is
expressed
as
\begin{equation}
{\rm i}\, \frac{d\nu}{dx} = H\nu ,     \label{shiki2}
\end{equation}
where $x$ is the time in which the neutrino propagates and
\begin{eqnarray}
H &\equiv& -U{\rm diag}(p_1, p_2, p_3, p_4)U^\dagger    \nonumber  \\
&\simeq & \frac{1}{2E}U{\rm diag}(\mu_1^2, \mu_2^2, \mu_3^2,
\mu_4^2)U^\dagger   \label{shiki3}
\end{eqnarray}
with a unitary mixing matrix $U$, energy of neutrino $E$, and the effective
mass squared $\mu_i^2$'s $(i = 1, 2, 3, 4)$. In Eq.(3) we have taken an
approximation that neutrino masses are much smaller than their momenta and
energies and have neglected an irrelevant term to the neutrino oscillation.
The matrix $U$ and the masses $\mu_i$'s are determined by
\begin{eqnarray}
U \left(
\begin{array}{@{\,}cccc@{\,}}
\mu_1^2 & 0 & 0 & 0 \\
0 & \mu_2^2 & 0 & 0 \\
0 & 0 & \mu_3^2 & 0 \\
0 & 0 & 0 & \mu_4^2
\end{array}
\right) U^\dagger &=& U^{(0)} \left(
\begin{array}{@{\,}cccc@{\,}}
0 & 0 & 0 & 0 \\
0 & \Delta m_{21}^2 & 0 & 0 \\
0 & 0 & \Delta m_{31}^2 & 0 \\
0 & 0 & 0 & \Delta m_{41}^2
\end{array}
\right) U^{(0)\dagger}   \nonumber  \\
&+& \left(
\begin{array}{@{\,}cccc@{\,}}
a & 0 & 0 & 0 \\
0 & 0 & 0 & 0 \\
0 & 0 & 0 & 0 \\
0 & 0 & 0 & a'
\end{array}
\right)  ,   \label{shiki4}
\end{eqnarray}
where $\Delta m_{ij}^2 = m_i^2-m_j^2$ and
\begin{eqnarray}
a &\equiv & 2{\sqrt 2}G_FN_eE = 7.60\times 10^{-5}\frac{\rho}
{[{\rm g}\,{\rm cm}^{-3}]}\frac{E}{[{\rm GeV}]}\;\; {\rm eV}^2 ,
\nonumber  \\
a' &\equiv & {\sqrt 2}G_FN_nE \simeq {\sqrt 2}G_FN_eE = a/2 .
\label{shiki5}
\end{eqnarray}
The quantities $a$ and $a'$ denote the matter effect to the oscillation, $a$
coming from the charged current process of $\nu_e$ and $a'$ from the neutral
current process of $\nu_e, \nu_\mu$, and $\nu_\tau$. In Eq.(5), $N_e$ is the
electron density of the matter, $\rho$ is the matter density, and $N_n$ is
the
neutron density which is approximately equal to $N_e$ since we consider the
earth matter effect in the long-baseline experiments.
The solution of Eq.(2) is given by
\begin{equation}
\nu (x) = S(x)\nu (0) ,    \label{shiki6}
\end{equation}
with
\begin{equation}
S(x) = T{\rm exp}\left( -{\rm i}\int_0^x ds H(s) \right)  ,   \label{shiki7}
\end{equation}
where $T$ is the time ordering operator, and $x$ is actually the distance in
which the neutrino propagates with the speed almost equal to the light
velocity. In the following, the matter density is assumed to be independent
of
space and time for simplicity, and then we have
\begin{equation}
S(x) = {\rm e}^{-{\rm i}Hx}  .  \label{shiki8}
\end {equation}
The oscillation probability for $\nu_\alpha\to\nu_\beta$ for the distance
$L$
is expressed as
\begin{equation}
P(\nu_\alpha\to\nu_\beta ; L) = \left| S_{\beta\alpha}(L) \right|^2 .
\label{shiki9}
\end{equation}
The oscillation probability for the antineutrinos $P(\bar{\nu_\alpha}\to
\bar{\nu_\beta} ; L)$ is obtained by replacing $U \to U^*, a \to -a$, and
$a' \to -a'$ in Eq.(9). The CP violation effect in the neutrino oscillation
is
given by the probability difference between CP-conjugate channels as
follows:
\begin{equation}
\Delta P(\nu_\alpha\to\nu_\beta)  \equiv P(\nu_\alpha\to\nu_\beta ; L) -
P(\bar{\nu_\alpha}\to\bar{\nu_\beta} ; L) .   \label{shiki10}
\end{equation}
This quantity $\Delta P(\nu_\alpha\to\nu_\beta)$ consists of the pure
CP-violation effect due to the phases of $U^{(0)}$ and the fake CP-violation
effect due to the matter effect.

In the four-neutrino model, the four neutrino masses can be divided into two
classes: 3+1 and 2+2 schemes. The 2+2 scheme consists of the two pairs of
close masses separated by the LSND mass gap of the order of 1 eV
\cite{BilenkyEPJ}\cite{Okada}\cite{Barger98} so as to accomodate
the solar and atmospheric neutrino deficits and the LSND experiments
together with the results from the other accelerator and reactor experiments
on the neutrino oscillation. The 3+1 scheme consists of a group of three
masses separated from an isolated one by the gap of the order of 1 eV. This
scheme is only marginally allowed\cite{Kayser} and the phenomenology
including
CP violation is discussed by Donini and Maloni \cite{Donini} together with
the
2+2 scheme, showing that the detailed comparison of the physical reach of
the
neutrino factory in the two schemes gives similar results for the
sensitivity
to the mixing angles. We concentrate here on the 2+2 scheme in ordr to see
the CP violation effect in the oscillation for various rates of the
active-sterile admixture of neutrinos, as stated in the Introduction.
There are the following two mass patterns in the 2+2 scheme;
(i) $\Delta m^2_{{\rm solar}} \equiv \Delta m^2_{21} \ll
\Delta m^2_{{\rm atm}} \equiv \Delta m^2_{43} \ll
\Delta m^2_{{\rm LSND}} \equiv \Delta m^2_{32}$, and (ii)
$\Delta m^2_{{\rm solar}} \equiv \Delta m^2_{43} \ll
\Delta m^2_{{\rm atm}} \equiv \Delta m^2_{21} \ll
\Delta m^2_{{\rm LSND}} \equiv \Delta m^2_{32}$. We will
adopt the first pattern in the following analyses, and the second pattern
can
be attained only through the exchange of indices $(1, 2) \leftrightarrow
(3, 4)$ in the following various expressions such as the oscillation
probabilities.

Since $\Delta m^2_{21} \ll \Delta m^2_{31}, \Delta m^2_{41}$ and $a, a' \ll
\Delta m^2_{31}, \Delta m^2_{41}$, we decompose $H$ as $H = H_0 + H_1$
of Eq.(3) with
\begin{equation}
H_0 = \frac{1}{2E}U^{(0)} \left (
\begin{array}{@{\,}cccc@{\,}}
0 & 0 & 0 & 0 \\
0 & 0 & 0 & 0 \\
0 & 0 & \Delta m_{31}^2 & 0 \\
0 & 0 & 0 & \Delta m_{41}^2
\end{array}
\right ) U^{(0)\dagger} ,        \label{shiki11}
\end{equation}
and
\begin{equation}
H_1 = \frac{1}{2E} U^{(0)} \left(
\begin{array}{@{\,}cccc@{\,}}
0 & 0 & 0 & 0 \\
0 & \Delta m_{21}^2 & 0 & 0 \\
0 & 0 & 0 & 0 \\
0 & 0 & 0 & 0
\end{array}
\right) U^{(0)\dagger} + \frac{1}{2E} \left(
\begin{array}{@{\,}cccc@{\,}}
a & 0 & 0 & 0 \\
0 & 0 & 0 & 0 \\
0 & 0 & 0 & 0 \\
0 & 0 & 0 & a'
\end{array}
\right)  ,     \label{shiki12}
\end{equation}
and treat $H_1$ as a perturbation and calculate Eq.(8) up to the first order
in $a, a'$, and $\Delta m_{21}^2$. Following the Arafune-Koike-Sato
procedure
\cite{Arafune}, $S(x)$ of Eq.(8) is given by
\begin{equation}
S(x) \simeq {\rm e}^{-{\rm i}H_0x} - {\rm i}\, {\rm e}^{-{\rm i}H_0x}
\int_0^x ds H_1(s) ,   \label{shiki13}
\end{equation}
where $H_1(x) = {\rm e}^{{\rm i}H_0x}H_1{\rm e}^{-{\rm i}H_0x}$.
The approximation in Eq.(13) requires
\begin{equation}
\frac{\Delta m_{21}^2L}{2E} \ll 1 , \qquad
\frac{aL}{2E} \ll 1 , \qquad \frac{a'L}{2E} \ll 1 .
\label{shiki14}
\end{equation}
The requirements of Eq.(14) are satisfied for
$\Delta m_{21}^2=(10^{-5}-10^{-4})$ ${\rm eV}^2$, $\Delta m_{31}^2=
(0.1-1)$ ${\rm eV}^2$, $E=1-15$ ${\rm GeV}$, $L = (300-750)$ km, and $\rho =
3
{\rm g/cm^3}$ as
\begin{equation}
\frac{\Delta m_{21}^2L}{2E} \simeq 5\times 10^{-4}-0.2 , \qquad
\frac{aL}{2E} , \;\; \frac{a'L}{2E} \simeq 0.1-0.4 .   \label{shiki15}
\end{equation}
Equation (14) also shows  that the approximation becomes better as the
energy
$E$ increases, so we can apply this approximation to the multi-GeV region
such
as $E = 1-15$ ${\rm GeV}$. If we express $S_{\beta\alpha}(x)$ as
\begin{equation}
S_{\beta\alpha}(x) = \delta_{\beta\alpha} + {\rm i}\, T_{\beta\alpha}(x) ,
\label{shiki16}
\end{equation}
then ${\rm i}T_{\beta\alpha}(x)$ is obtained as follows (in the following,
$U^{(0)}_{\beta\alpha}$ is denoted as $U_{\beta\alpha}$ for brevity):
\begin{eqnarray}
{\rm i}T_{\beta\alpha}(x) &=& -2\, {\rm i}\, {\rm exp}\left( -{\rm i}
\frac{\Delta m^2_{31}x}{4E} \right)
\sin\left( \frac{\Delta m^2_{31}x}{4E} \right)
[\> U_{\beta 3}U_{\alpha 3}^* \{ 1-\frac{a}{\Delta m^2_{31}}
(2|U_{e3}|^2    \nonumber  \\
& & {} -\delta_{\alpha e}-\delta_{\beta e})
-\frac{a'}{\Delta m^2_{31}}(2|U_{s3}|^2-\delta_{\alpha s}
-\delta_{\beta s})-{\rm i}\,\frac{ax}{2E}|U_{e3}|^2    \nonumber  \\
& & {} -{\rm i}\,\frac{a'x}{2E}|U_{s3}|^2 \}
-( \frac{a}{\Delta m^2_{31}}+\frac{a}{\Delta m^2_{43}} )
(U_{\alpha 3}^*U_{\beta 4}U_{e3}U_{e4}^*
+U_{\alpha 4}^*U_{\beta 3}U_{e4}U_{e3}^*)  \nonumber  \\
& & {} -( \frac{a'}{\Delta m^2_{31}}+\frac{a'}{\Delta m^2_{43}} )
(U_{\alpha 3}^*U_{\beta 4}U_{s3}U_{s4}^*
+U_{\alpha 4}^*U_{\beta 3}U_{s4}U_{s3}^*)\> ]  \nonumber  \\
& & {} -2\, {\rm i}\, {\rm exp}\left( -{\rm i}\frac{\Delta m^2_{41}x}{4E}
\right) \sin\left( \frac{\Delta m^2_{41}x}{4E}\right) [\> U_{\beta 4}
U_{\alpha 4}^*\{ 1-\frac{a}{\Delta m^2_{41}}(2|U_{e4}|^2
\nonumber  \\
& & {} -\delta_{\alpha e}-\delta_{\beta e})
-\frac{a'}{\Delta m^2_{41}}(2|U_{s4}|^2-\delta_{\alpha s}
-\delta_{\beta s})-{\rm i}\,\frac{ax}{2E}|U_{e4}|^2    \nonumber  \\
& & {} -{\rm i}\,\frac{a'x}{2E}|U_{s4}|^2 \}
-( \frac{a}{\Delta m^2_{41}}-\frac{a}{\Delta m^2_{43}} )
(U_{\alpha 3}^*U_{\beta 4}U_{e3}U_{e4}^*
+U_{\alpha 4}^*U_{\beta 3}U_{e4}U_{e3}^*)  \nonumber  \\
& & {} -( \frac{a'}{\Delta m^2_{41}}-\frac{a'}{\Delta m^2_{43}} )
(U_{\alpha 3}^*U_{\beta 4}U_{s3}U_{s4}^*
+U_{\alpha 4}^*U_{\beta 3}U_{s4}U_{s3}^*)\> ]  \nonumber  \\
& & {} -{\rm i}\,\frac{\Delta m^2_{31}x}{2E}[\>\frac{\Delta m_{21}^2}
{\Delta m_{31}^2}U_{\beta 2}U_{\alpha 2}^*+\frac{a}{\Delta m^2_{31}}
\{ \delta_{\alpha e}\delta_{\beta e}+U_{\beta 3}U_{\alpha 3}^*
(2|U_{e3}|^2-\delta_{\alpha e}    \nonumber  \\
& & {} -\delta_{\beta e})+U_{\beta 4}U_{\alpha 4}^*(2|U_{e4}|^2
-\delta_{\alpha e}-\delta_{\beta e})+U_{\alpha 3}^*U_{\beta 4}
U_{e3}U_{e4}^*    \nonumber  \\
& & {} +U_{\alpha 4}^*U_{\beta 3}U_{e4}U_{e3}^* \}
+\frac{a'}{\Delta m^2_{31}}\{ \delta_{\alpha s}\delta_{\beta s}
+U_{\beta 3}U_{\alpha 3}^*(2|U_{s3}|^2-\delta_{\alpha s}
-\delta_{\beta s})    \nonumber  \\
& & {} +U_{\beta 4}U_{\alpha 4}^*(2|U_{s4}|^2-\delta_{\alpha s}
-\delta_{\beta s})+U_{\alpha 3}^*U_{\beta 4}U_{s3}U_{s4}^*
\nonumber  \\
& & {} +U_{\alpha 4}^*U_{\beta 3}U_{s4}U_{s3}^* \}\> ] .
\label{shiki17}
\end{eqnarray}
We use Eq.(17) in Eq.(16) and calculate the oscillation probability for
$\nu_\alpha\to\nu_\beta$ by Eq.(9). The complete expression of
$P(\nu_\alpha\to\nu_\beta; L)$ in the four-neutrino model with matter effect
is given in the Appendix.
\vskip 0.7truecm

\centerline{\large\bf III  Constraints on the mixing matrix}
\vskip 0.2truecm

In this section the constraints imposed on the mixing matrix $U$ are derived
from the solar neutrino deficit, atmospheric neutrino anomaly, LSND
experiments and the other terrestrial oscillation experiments using the
accelerators and reactors.

\noindent
(i) We use the results of the recent combined analysis of the atmospheric
neutrino anomaly and the solar neutrino deficit in the four-neutrino scheme,
done by Gonzalez-Garcia, Maltoni, and Pe\~na \cite{Gonzalez-Garcia}. They
obtained two solutions; (A) close-to-active solar neutrino oscillations plus
close-to-sterile atmospheric neutrino oscillations, expressed by
\begin{equation}
|U_{s1}|^2+|U_{s2}|^2 \sim 0.2 ,   \label{shiki18}
\end{equation}
and (B) near-pure-sterile solar neutrino oscillations plus near-pure-active
atmospheric neutrino oscillations, expressed by
\begin{equation}
|U_{s1}|^2+|U_{s2}|^2 \sim 0.91-0.97 .   \label{shiki19}
\end{equation}
For the later convenience, we define the quantity $|U_{s1}|^2+|U_{s2}|^2$ as
$D$.

\noindent
(ii) A constraint on $U_{\mu 3}$ and $U_{\mu 4}$ is derived from
the atmospheric neutrino anomaly, where the survival probability of
$\nu_\mu$
is given by
\begin{equation}
P(\nu_{\mu}\to\nu_{\mu}) \simeq 1 - 4|U_{\mu 3}|^2
|U_{\mu 4}|^2\sin^2\Delta_{43} - 2(|U_{\mu 1}|^2+|U_{\mu 2}|^2)
(1-|U_{\mu 1}|^2-|U_{\mu 2}|^2) ,   \label{shiki20}
\end{equation}
where $\Delta_{ij}\equiv \Delta m^2_{ij}L/(4E)$. The Super-Kamiokande
data, $\sin^22\theta_{{\rm atm}}>0.82$ for $5\times 10^{-4}<
\Delta m^2_{{\rm atm}}<6\times 10^{-3}$ ${\rm eV}^2$ \cite{Fukuda98},
gives a constraint, along with the expectation of
$|U_{\mu 1}|^2+|U_{\mu 2}|^2\ll 1$, of
\begin{equation}
|U_{\mu 3}|^2|U_{\mu 4}|^2 > 0.205 .  \label{shiki21}
\end{equation}
(iii) The Bugey experiment of short-baseline reactor $\bar{\nu_e}$
disappearance \cite{Bugey} gives a constraint on $|U_{e3}|^2+|U_{e4}|^2$.
The survival probability of $\bar{\nu_e}$ is expressed by
\begin{equation}
P(\bar{\nu_e}\to\bar{\nu_e}) \simeq 1 - 4(|U_{e3}|^2
+|U_{e4}|^2)(1-|U_{e3}|^2-|U_{e4}|^2)\sin^2\Delta_{32} ,  \label{shiki22}
\end{equation}
where $\Delta_{41}\sim \Delta_{42}\sim \Delta_{31}\sim \Delta_{32}$
is used. The data, $\sin^22\theta_{{\rm Bugey}}<0.1$ for $0.1<\Delta m^2<
1$ ${\rm eV}^2$, brings a constraint of
\begin{equation}
|U_{e3}|^2+|U_{e4}|^2 < 0.025 .  \label{shiki23}
\end{equation}
The first long-baseline reactor experiment, that is, the CHOOZ
experiment \cite{CHOOZ} gives a constraint of $4|U_{e3}|^2|U_{e4}|^2 <
0.18$,
through their data of $\sin^22\theta_{{\rm CHOOZ}}<0.18$ for
$3\times 10^{-3}<\Delta m^2<1.0\times 10^{-2}$ ${\rm eV}^2$. This
constraint can be involved in the constraint of Eq.(23) obtained from the
Bugey experiment.

\noindent
(iv) In the same way as above, the LSND experiment \cite{LSND} brings
a constraint of
\begin{equation}
|U^*_{\mu 3}U_{e3} +U^*_{\mu 4}U_{e4}| = 0.016 - 0.12    \label{shiki24}
\end{equation}
from the data of $\sin^22\theta_{\rm LSND}=1.0\times 10^{-3}-6.0
\times 10^{-2}$ for $0.2<\Delta m^2_{\rm LSND}<2$ ${\rm eV}^2$.

\noindent
(v) The CHORUS \cite{CHORUS} and NOMAD \cite{NOMAD} experiments
searching for the $\nu_{\mu}\to\nu_{\tau}$ oscillation gives a constraint of
\begin{equation}
|U^*_{\mu 3}U_{\tau 3} +U^*_{\mu 4}U_{\tau 4}| < 0.4    \label{shiki25}
\end{equation}
for $\Delta m^2<1$ ${\rm eV^2}$, which is derived from the latest NOMAD
experimental data of $\sin^22\theta_{{\rm NOMAD}} < (0.8{\rm eV^2}/
\Delta m^2)^2$ for $0.8<\Delta m^2<10$ ${\rm eV}^2$.

The details of the derivation of the constraints in (ii)-(v) can be seen in
our previous
work \cite{Hattori}.
\vskip 0.7truecm

\centerline{\large\bf IV. CP violation and matter effect}
\vskip 0.2truecm

In this section we will investigate how large the CP violation effect could
be
in the long-baseline neutrino oscillations in the light of the recent
combined
analysis of the solar and atmospheric neutrino deficits, that is, depending
on
the rate of the active-sterile neutrino admixture, and how much the matter
effect affects the pure CP violation effect.

In order to translate the constraints on $U$ derived in the previous section
into the ones with the mixing angles
and phases, we adopt the most general parametrization of $U$ for Majorana
neutrinos \cite{Barger}, which includes six mixing angles and six phases.
The expression of the matrix is too complicated to show it here, so that
we cite only the matrix elements which are useful for the following
analyses;
$U_{e1}=c_{01}c_{02}c_{03}, U_{e2}=c_{02}c_{03}s^*_{d01},
U_{e3}=c_{03}s^*_{d02}, U_{e4}=s^*_{d03},
U_{\mu 3}=-s^*_{d02}s_{d03}s^*_{d13}+c_{02}c_{13}s^*_{d12},
U_{\mu 4}=c_{03}s^*_{d13}, U_{\tau 3}=-c_{13}s^*_{d02}s_{d03}
s^*_{d23}-c_{02}s^*_{d12}s_{d13}s^*_{d23}+c_{02}c_{12}c_{23},
U_{\tau 4}=c_{03}c_{13}s^*_{d23}, U_{s3}=-c_{13}s^*_{d02}s_{d03}
c_{23}-c_{02}s^*_{d12}s_{d13}c_{23}-c_{02}c_{12}s_{d23}$, and
$U_{s4}=c_{03}c_{13}c_{23}$ (instead of $U_{s1}$ and $U_{s2}$) ,
where $c_{ij}\equiv \cos\theta_{ij}$
and $s_{dij}\equiv s_{ij}{\rm e}^{{\rm i}\delta_{ij}}\equiv \sin\theta_{ij}
{\rm e}^{{\rm i}\delta_{ij}}$, and $\theta_{01}, \theta_{02},
\theta_{03}, \theta_{12}, \theta_{13}, \theta_{23}$ are the six angles and
$\delta_{01}, \delta_{02}, \delta_{03}, \delta_{12}, \delta_{13},
\delta_{23}$ are the six phases. Three of the six oscillation
probability differences are independent so that only three of the six phases
are determined by the measurements of the CP violation effects, that is,
the Dirac phases.

By using this parametrization of $U$, the constraints of Eqs.(18), (19),
(21),
(23), (24) and (25) are expressed by the mixing angles and phases as
follows:
\begin{eqnarray}
&|&s_{02}s_{03}c_{13}c_{23}{\rm e}^{-{\rm
i}\delta_1}+c_{02}c_{23}s_{12}s_{13}
+c_{02}c_{12}s_{23}{\rm e}^{{\rm i}\delta_2}|^2+|c_{03}c_{13}c_{23}|^2
\nonumber  \\
& & {} \sim  0.8\;\; {\rm (A)}\qquad {\rm or}\qquad 0.03-0.09\;\;{\rm (B)} ,
\label{shiki26}
\end{eqnarray}
\begin{equation}
 | -s_{02}s_{03}s_{13}{\rm e}^{-{\rm i}\delta_1}+c_{02}c_{13}s_{12} |^2
c^2_{03}s^2_{13} > 0.205 ,   \label{shiki27}
\end{equation}
\begin{equation}
c^2_{03}s^2_{02} + s^2_{03} < 0.025 .  \label{shiki28}
\end{equation}
\begin{equation}
 |\> c_{02}s_{02}c_{03}s_{12}c_{13} + c^2_{02}c_{03}s_{03}s_{13}
{\rm e}^{{\rm i}\delta_1}| = 0.016 - 0.12 ,  \label{shiki29}
\end{equation}
\begin{eqnarray}
&|&c^2_{02}c_{12}s_{12}c_{13}c_{23}-c_{02}s_{02}s_{03}s_{12}
c^2_{13}s_{23}{\rm e}^{-{\rm i}(\delta_1+\delta_2)}-c_{02}s_{02}
s_{03}c_{12}s_{13}c_{23}{\rm e}^{{\rm i}\delta_1}   \nonumber  \\
& & {} +c_{02}s_{02}s_{03}s_{12}s^2_{13}s_{23}{\rm e}^{{\rm i}(\delta_1
-\delta_2)}+c_{13}s_{13}s_{23}(c^2_{03}-c^2_{02}s^2_{12}
+s^2_{02}s^2_{03}){\rm e}^{-{\rm i}\delta_2}|   \nonumber  \\
& & {} < 0.4 ,   \label{shiki30}
\end{eqnarray}
where $\delta_1\equiv \delta_{02}-\delta_{03}-\delta_{12}+\delta_{13}$
and $\delta_2\equiv \delta_{12}-\delta_{13}+\delta_{23}$. The constraint of
Eq.(26) is expressed for $|U_{s3}|^2+|U_{s4}|^2$ instead of the one for
$|U_{s1}|^2+|U_{s2}|^2$ in Eqs.(18) and (19).

Equation (28) from the Bugey experiment gives a stringent constraint on the
two mixing angles $s_{02}$ and $s_{03}$, like the one for $|U_{e3}|$ from
the
CHOOZ experiment in the three-neutrino mixing scheme. And, in this
situation,
Eq.(27) from the atmospheric neutrinos gives a strong constraint on the
mixing angles $s_{12}$ and $s_{13}$, roughly $s_{12}>0.91$ and $s_{13}$
around
the maximal mixing. The angle $s_{23}$ strongly affects the rate of the
active-sterile admixture, $|U_{s1}|^2+|U_{s2}|^2$ $(\equiv D)$, as seen in
Eq.(26). The angle $s_{01}$ does not occur in Eqs.(26)-(30). The phase
$\delta_1$ affects the the determination of the allowed regions of $s_{02},
s_{03}, s_{12}$, and $s_{13}$ through Eqs.(27) and (29).

In order to see the gross features of the pure CP violation effect in the
long-baseline $\nu_\mu\to\nu_e$ and $\nu_\mu\to\nu_\tau$ oscillations with
respect to the mixing angles and phases, we write down the expressions of
the
effect to the leading terms relevant to the long-baseline oscillation and by
using the smallness of $s_{02}$ and $s_{03}$ as follows:
\begin{eqnarray}
\Delta P(\nu_\mu\to\nu_e)&\simeq & 4c_{02}s_{02}c^2_{03}s_{03}s_{12}c_{13}
s_{13}\sin\delta_1\sin\left( \frac{\Delta m^2_{43}L}{2E} \right) ,
\label{shiki31}  \\
\Delta P(\nu_\mu\to\nu_\tau)&\simeq & -4c^2_{02}c^2_{03}c_{12}s_{12}c^2_{13}
s_{13}c_{23}s_{23}\sin\delta_2\sin\left( \frac{\Delta m^2_{43}L}{2E} \right)
.
\label{shiki32}
\end{eqnarray}
These expressions are obtained from the exact expression of the pure CP
violation effect of Eq.(10), not from the approximate one given in the
Appendix. As can be seen in Eqs.(31) and (32), $\Delta P(\nu_\mu\to\nu_e)$
depends primarily on the phase $\delta_1$ and $\Delta P(\nu_\mu\to\nu_\tau)$
depends on the phase $\delta_2$. The angle $s_{23}$ determines
$\Delta P(\nu_\mu\to\nu_\tau)$, but does not affect $\Delta P(\nu_\mu
\to\nu_e)$ to the leading terms. Similarly, the angles $s_{02}$ and $s_{03}$
determine $\Delta P(\nu_\mu\to\nu_e)$, but do not appreciably affect
$\Delta P(\nu_\mu\to\nu_\tau)$, since $s_{02}$ and $s_{03}$ are very small.
So, first, in Fig.1 we show the pure CP violation effect in
$\nu_\mu\to\nu_e$
oscillation as a function of the phase $\delta_1$ for the baseline of
$L = 290$ km and the neutrino energy $E = 1.2$ ${\rm GeV}$ for the typical
three parameter sets which are allowed by the constraints of Eqs.(26)-(30);
$(s_{02} = 0.12, s_{03} = 0.06, s_{12} = 0.93, s_{13} = 0.71), (s_{02} =
0.12,
s_{03} = 0.05, s_{12} = 0.97, s_{13} = 0.71)$, and $(s_{02} = 0.15, s_{03} =
0.02, s_{12} = 0.95, s_{13} = 0.71)$, and commonly $s_{01} = s_{23} =
1/\sqrt{2}$ and $\delta_2 = \pi /2$. We have taken the mass-squared
differences as $\Delta m^2_{43} = \Delta m^2_{\rm atm} = 2.5\times 10^{-3}
{\rm eV}^2, \Delta m^2_{32} = \Delta m^2_{\rm LSND} = 0.3{\rm eV}^2$, and
$\Delta m^2_{21} = \Delta m^2_{\rm solar} = 5\times 10^{-5}{\rm eV}^2$,
which
are fixed in the following unless stated otherwise. As seen in Fig.1, the
magnitude of CP violation in $\nu_\mu\to\nu_e$ oscillation is at most 0.03
for
$L = 290$ km and at $E = 1.2$ ${\rm GeV}$, which is almost the same
in magnitude as in the three-neutrino mixing scheme. In Fig.2 we show the
pure
CP violation effect in $\nu_\mu\to\nu_\tau$ oscillation as a function of the
phase $\delta_2$ at $L = 730$ km and $E = 6.1$ ${\rm GeV}$ for the typical
three parameter sets; $(s_{02} = 0.12, s_{03} = 0.06, s_{12} = 0.93, s_{13}
=
0.71), (s_{02} = 0.12, s_{03} = 0.06, s_{12} = 0.97, s_{13} = 0.71)$, and
$(s_{02} = 0.15, s_{03} = 0.03, s_{12} = 0.99, s_{13} = 0.71)$, and commonly
$s_{01} = s_{23} = 1/\sqrt{2}$ and $\delta_1 = \pi /2$. As seen in Fig.2,
the
magnitude of CP violation effect in $\nu_\mu\to\nu_\tau$ oscillation could
attain as large as 0.3, as already shown in Ref. 35. We show in Fig.3 the
pure
CP violation effect in $\nu_\mu\to\nu_\tau$ oscillation as a function of the
mixing angle $\theta_{23}$ at $L = 730$ km and $E = 6.1$ ${\rm GeV}$ for the
three parameter sets; $(s_{02} = s_{03} = 0.11, s_{12} = 0.93, s_{13} =
0.71),
(s_{02} = 0.15, s_{03} = 0.05, s_{12} = 0.95, s_{13} = 0.71)$, and
$(s_{02} = s_{03} = 0.11, s_{12} = 0.97, s_{13} = 0.71)$, and commonly
$s_{01} = 1/\sqrt{2}$ and $\delta_1 = \delta_2 = \pi /2$. As seen in Fig.3,
the maximal mixing of $\theta_{23} (\simeq 45^\circ)$ gives the maximum CP
violation
effect. As can be seen fron Eqs.(31)and (32), the CP violation effects
$\Delta P(\nu_\mu\to\nu_e)$ and $\Delta P(\nu_\mu\to\nu_\tau)$ do not depend
on the mixing angle $s_{01}$ to the leading terms. So, in the following
calculations we fix as $s_{01} = 1/\sqrt{2}$.

Next, we discuss the relation between the rate of active-sterile admixture
$|U_{s1}|^2+|U_{s2}|^2$ $(\equiv D)$ and the pure CP violation effect in
$\nu_\mu\to\nu_e$ and $\nu_\mu\to\nu_\tau$ oscillations, where $D$ is given
by
Eq.(26) subtracted from 1. As can be seen from Eq.(31), since the pure CP
violation effect $\Delta P(\nu_\mu\to\nu_e)$ does not depend on the mixing
angle $s_{23}$ to the leading terms, it does not vary with the quantity $D$.
So, $\Delta P(\nu_\mu\to\nu_e)$ is almost the same at the level of $\le
0.05$
at $L = 290$ km and in $1 \le E \le 10$ ${\rm GeV}$ among the
close-to-active
solar neutrino oscillations plus close-to-sterile atmospheric neutrino
oscillations
$(D \sim 0.2)$, near-pure-sterile solar neutrino oscillations plus
near-pure-active atmospheric neutrino oscillations $(D \sim 0.91-0.97)$, and
the maximal active-sterile admixture $(D \sim 0.5)$. We show in Fig.4 the
pure
CP violation effect of $\Delta P(\nu_\mu\to\nu_e)$ at $L = 290$ km in the
neutrino energy range of $1.5 \le E \le 3$ ${\rm GeV}$ for the two cases of
mixing angles; $(s_{02} = s_{03} = 0.11, s_{12} = 0.97, s_{13} = 0.73)$ in
solid line and $(s_{02} = 0.12, s_{03} = 0.06, s_{12} = 0.93, s_{13} =
0.71)$
in dashed line, for commonly $s_{23} = 0.4$ and $\delta_1 = \delta_2 = \pi
/2$.
In order to see the magnitude of the matter effect to $\Delta P(\nu_\mu
\to\nu_e)$, we show in Figs.5 and 6 the pure CP violation effect(solid line)
and the matter effect(dotted line) for the above-mentioned two cases of the
mixing angles, respectively. The matter effect is calculated from the
equation in the Appendix and, as can be seen in Figs.5 and 6, its relative 
magnitude to the pure CP violation effect is around 15\% at 1.5GeV and 
30\% at 3GeV. Fig.7 shows the pure CP violation effect and the matter effect 
in $\nu_\mu\to\nu_e$
oscillation in the high energy range of $3 \le E \le 10$ ${\rm GeV}$ for
$(s_{02} = s_{03} = 0.11, s_{12} = 0.97, s_{13} = 0.73)$. The relative
magnitude of the matter effect becomes larger than that in the energy range
of $1.5 \le E \le 3$ ${\rm GeV}$.

\begin{table}
\caption{The pure CP violation effect in $\nu_\mu\to\nu_\tau$ oscillation,
$\Delta P(\nu_\mu\to\nu_\tau)$, at the baseline $L=730$ km and neutrino
energy
$E=6.1$ GeV, and the active-sterile admixture $D$ with respect to the mixing
angle $s_{23}$ and the phase $\delta_2$ for the parameter set of $(s_{02}
=0.12, s_{03}=0.06, s_{12}=0.93, s_{13}=0.71)$ and $\delta_1=\pi /2$.}
\begin{center}
\begin{tabular}{|c|c|c|c|c|c|c|}  \hline
\multicolumn{1}{|c|}{ } & \multicolumn{2}{|c|}{$\delta_2=30^\circ$} &
\multicolumn{2}{|c|}{$\delta_2=60^\circ$} & \multicolumn{2}{|c|}{$\delta_2
=90^\circ$}  \\  \hline
$s_{23}$ & $D$ & $\Delta P(\nu_\mu\to\nu_\tau)$ & $D$ & $\Delta P(\nu_\mu
\to\nu_\tau)$ & $D$ & $\Delta P(\nu_\mu\to\nu_\tau)$  \\  \hline
0.00 & 0.08 &  0.00 & 0.08 &  0.00 & 0.08 &  0.00  \\
0.10 & 0.04 & -0.03 & 0.06 & -0.06 & 0.08 & -0.06  \\
0.20 & 0.03 & -0.06 & 0.06 & -0.11 & 0.11 & -0.13  \\
0.30 & 0.03 & -0.09 & 0.08 & -0.16 & 0.15 & -0.18  \\
0.40 & 0.05 & -0.12 & 0.12 & -0.20 & 0.20 & -0.24  \\
0.50 & 0.10 & -0.14 & 0.17 & -0.24 & 0.28 & -0.28  \\
0.60 & 0.16 & -0.16 & 0.25 & -0.27 & 0.36 & -0.31  \\
0.65 & 0.21 & -0.16 & 0.29 & -0.28 & 0.41 & -0.32  \\
0.70 & 0.26 & -0.17 & 0.35 & -0.28 & 0.47 & -0.33  \\
0.75 & 0.32 & -0.17 & 0.40 & -0.28 & 0.52 & -0.32  \\
0.80 & 0.38 & -0.16 & 0.47 & -0.28 & 0.58 & -0.32  \\
0.85 & 0.46 & -0.15 & 0.54 & -0.26 & 0.65 & -0.30  \\
0.90 & 0.55 & -0.14 & 0.62 & -0.23 & 0.72 & -0.26  \\
0.95 & 0.67 & -0.11 & 0.72 & -0.18 & 0.79 & -0.21  \\
1.00 & 0.87 & -0.02 & 0.87 & -0.02 & 0.87 & -0.02  \\  \hline
\end{tabular}
\end{center}
\label{tab1}
\end{table}
\vskip 0.1truecm

\begin{table}
\caption{The pure CP violation effect in $\nu_\mu\to\nu_\tau$ oscillation,
$\Delta P(\nu_\mu\to\nu_\tau)$, at $L=730$ km and $E=6.1$ GeV, and the
active-sterile admixture $D$ with respect to the mixing angle $s_{23}$ and
the phase $\delta_2$ for the parameter set of $(s_{02}=0.12, s_{03}=0.06,
s_{12}=0.97, s_{13}=0.72)$ and $\delta_1=\pi /2$.}
\begin{center}
\begin{tabular}{|c|c|c|c|c|c|c|}  \hline
\multicolumn{1}{|c|}{ } & \multicolumn{2}{|c|}{$\delta_2=30^\circ$} &
\multicolumn{2}{|c|}{$\delta_2=60^\circ$} & \multicolumn{2}{|c|}{$\delta_2
=90^\circ$}  \\  \hline
$s_{23}$ & $D$ & $\Delta P(\nu_\mu\to\nu_\tau)$ & $D$ & $\Delta P(\nu_\mu
\to\nu_\tau)$ & $D$ & $\Delta P(\nu_\mu\to\nu_\tau)$  \\  \hline
0.00 & 0.04 &  0.00 & 0.04 &  0.00 & 0.04 &  0.00  \\
0.10 & 0.02 & -0.02 & 0.03 & -0.04 & 0.05 & -0.04  \\
0.20 & 0.02 & -0.04 & 0.04 & -0.07 & 0.08 & -0.09  \\
0.30 & 0.04 & -0.06 & 0.07 & -0.11 & 0.12 & -0.13  \\
0.40 & 0.08 & -0.08 & 0.12 & -0.14 & 0.18 & -0.16  \\
0.50 & 0.14 & -0.10 & 0.19 & -0.17 & 0.27 & -0.19  \\
0.60 & 0.23 & -0.11 & 0.28 & -0.19 & 0.37 & -0.21  \\
0.65 & 0.28 & -0.11 & 0.34 & -0.19 & 0.42 & -0.22  \\
0.70 & 0.34 & -0.12 & 0.40 & -0.20 & 0.48 & -0.23  \\
0.75 & 0.40 & -0.12 & 0.46 & -0.20 & 0.55 & -0.23  \\
0.80 & 0.48 & -0.11 & 0.54 & -0.19 & 0.62 & -0.22  \\
0.85 & 0.56 & -0.11 & 0.62 & -0.18 & 0.69 & -0.21  \\
0.90 & 0.66 & -0.10 & 0.71 & -0.16 & 0.77 & -0.19  \\
0.95 & 0.77 & -0.08 & 0.80 & -0.13 & 0.85 & -0.15  \\
1.00 & 0.94 & -0.02 & 0.94 & -0.02 & 0.94 & -0.02  \\  \hline
\end{tabular}
\end{center}
\label{tab2}
\end{table}
\vskip 0.1truecm

\begin{table}
\caption{The pure CP violation effect in $\nu_\mu\to\nu_\tau$ oscillation,
$\Delta P(\nu_\mu\to\nu_\tau)$, at $L=730$ km and $E=6.1$ GeV, and the
active-sterile admixture $D$ with respect to the mixing angle $s_{23}$ and
the phase $\delta_2$ for the parameter set of $(s_{02}=0.12, s_{03}=0.06,
s_{12}=0.99, s_{13}=0.73)$ and $\delta_1=\pi /2$.}
\begin{center}
\begin{tabular}{|c|c|c|c|c|c|c|}  \hline
\multicolumn{1}{|c|}{ } & \multicolumn{2}{|c|}{$\delta_2=30^\circ$} &
\multicolumn{2}{|c|}{$\delta_2=60^\circ$} & \multicolumn{2}{|c|}{$\delta_2
=90^\circ$}  \\  \hline
$s_{23}$ & $D$ & $\Delta P(\nu_\mu\to\nu_\tau)$ & $D$ & $\Delta P(\nu_\mu
\to\nu_\tau)$ & $D$ & $\Delta P(\nu_\mu\to\nu_\tau)$  \\  \hline
0.00 & 0.02 &  0.00 & 0.02 &  0.00 & 0.02 &  0.00  \\
0.10 & 0.01 & -0.013 & 0.02 & -0.02 & 0.03 & -0.03  \\
0.20 & 0.02 & -0.025 & 0.04 & -0.04 & 0.06 & -0.05  \\
0.30 & 0.06 & -0.038 & 0.08 & -0.06 & 0.11 & -0.07  \\
0.40 & 0.11 & -0.049 & 0.14 & -0.08 & 0.17 & -0.10  \\
0.50 & 0.18 & -0.059 & 0.22 & -0.10 & 0.26 & -0.11  \\
0.60 & 0.28 & -0.067 & 0.32 & -0.11 & 0.37 & -0.13  \\
0.65 & 0.34 & -0.070 & 0.38 & -0.12 & 0.43 & -0.13  \\
0.70 & 0.40 & -0.072 & 0.44 & -0.12 & 0.49 & -0.14  \\
0.75 & 0.47 & -0.072 & 0.51 & -0.12 & 0.56 & -0.14  \\
0.80 & 0.55 & -0.072 & 0.59 & -0.12 & 0.64 & -0.13  \\
0.85 & 0.64 & -0.069 & 0.67 & -0.11 & 0.71 & -0.13  \\
0.90 & 0.73 & -0.064 & 0.76 & -0.10 & 0.80 & -0.11  \\
0.95 & 0.84 & -0.054 & 0.86 & -0.08 & 0.89 & -0.09  \\
1.00 & 0.98 & -0.019 & 0.98 & -0.02 & 0.98 & -0.02  \\  \hline
\end{tabular}
\end{center}
\label{tab3}
\end{table}
\vskip 0.1truecm

For $\nu_\mu\to\nu_\tau$ oscillation, the pure CP violation effect
$\Delta P(\nu_\mu\to\nu_\tau)$ depends on $s_{23}$ as can be
seen from Eq.(32) and therefore on the quantity $D$. In Tables 1-3, we
show $\Delta P(\nu_\mu\to\nu_\tau)$ as a function of the mixing angle
$s_{23}$
and the phase $\delta_2$, both of which largely affect the magnitude of $D$,
at $L = 730$ km and $E = 6.1$ GeV for the typical three cases of $(s_{02},
s_{03}, s_{12}, s_{13})$. The case of $(s_{02} = 0.12, s_{03} = 0.06, s_{12}
=
0.93, s_{13} = 0.71)$ and $\delta_2 = 90^\circ$ gives a class of the
possibly
maximum values of $\Delta P(\nu_\mu\to\nu_\tau)$ at that baseline and
neutrino
energy in the region of the mixing angles and phases allowed by
Eqs.(26)-(30).
In Figs.8-12 we show the relation between the rate of active-sterile
admixture
$D$ and the behavior of the CP violation effect in the $\nu_\mu\to\nu_\tau$
oscillation at $L = 730$ km in the energy range $6 \le E \le 15$ ${\rm
GeV}$.
In the case of the solution of the close-to-active solar neutrino
oscillations
plus close-to-sterile atmospheric neutrino oscillations $(D \sim 0.2)$ to
the
combined analysis of the solar and atmospheric neutrinos
\cite{Gonzalez-Garcia}, the pure CP violation effect of $\Delta P(\nu_\mu
\to\nu_\tau)$ is shown in Fig.8 for the two cases of $(s_{02} = 0.12, s_{03}
=
0.06, s_{12} = 0.93, s_{13} = 0.71, s_{23} = 0.40)$ and $(s_{02} = 0.12,
s_{03} = 0.06, s_{12} = 0.97, s_{13} = 0.72, s_{23} = 0.40)$, along with
the matter
effect for the first case of the above two cases. The first case represents
the possibly maximum value of the pure $\Delta P(\nu_\mu\to\nu_\tau)$ for
this
solution $(D \sim 0.2)$. The matter effect is around 8\% at 6 GeV and 15\%
at
10 GeV relative to the pure CP violation effects, and is about half the one
of
the $\nu_\mu\to\nu_e$ oscillation. Fig.9 shows the pure CP violation
effect(solid line) and the magnitude of the matter effect(dotted line) for
the
second case of the above two cases. The matter effect is very small in $6
\le
E \le 12$ ${\rm GeV}$ in comparison with the pure CP violation effect. We
show
in Fig.10 the pure $\Delta P(\nu_\mu\to\nu_\tau)$ and the matter effect for
the solution of the near-pure-sterile solar neutrino oscillations plus
near-pure-active atmospheric neutrino oscillations $(D \sim 0.91-0.97)$ for
the typical case of $(s_{02} = 0.12, s_{03} = 0.06, s_{12} = 0.95, s_{13} =
0.71, s_{23} = 1.0)$. The pure CP violation effect is very small,
$\sim -0.01$, and is comparable to the matter effect.

Incidentally, we show in Fig.11 the pure $\Delta P(\nu_\mu\to\nu_\tau)$ and
the matter effect for the case of the maximal active-sterile
admixture $(D \sim 0.5)$, which is not allowed by the combined analysis of
the
solar and atmospheric neutrinos \cite{Gonzalez-Garcia}, for the two
parameter
sets of $(s_{02} = 0.12, s_{03} = 0.06, s_{12} = 0.93, s_{13} = 0.71,
s_{23} = 0.75)$ and $(s_{02} = 0.12, s_{03} = 0.06, s_{12} = 0.97,
s_{13} = 0.73, s_{23} = 0.70)$, along with the matter effect for the first
set. The first parameter set represents the possibly maximum value of the
pure $\Delta P(\nu_\mu\to\nu_\tau)$ in this case. Fig.12 shows the pure CP
violation effect(solid line) and the matter effect(dotted line) for the
second set of the above two sets. The matter effect is very small in
$6 \le E \le 15$ ${\rm GeV}$ in comparison with the pure CP violation
effect.
\vskip 0.7truecm

\centerline{\large\bf V  Conclusion}
\vskip 0.2truecm

We have evaluated the pure CP violation effect and the fake one due to
the matter effect in the long-baseline neutrino oscillations for the
baselines
$L=290$km and 730km in the neutrino energy range $E = 1.5-15$ GeV in the
four-neutrino model with the 2+2 scheme, where two pairs of two close
neutrino
masses are separated by the LSND mass gap of the order of 1eV, on the basis
of
the constraints on the mixing matrix from the solar neutrino deficit,
atmospheric neutrino anomaly, LSND experiments, Bugey and CHOOZ
measurements,
and CHORUS and NOMAD experiments. The matter effect is estimated with
Arafune-Koike-Sato's approximation method\cite{Arafune}. The matter effect
is
at $(15-30)\%$ level, relative to the pure CP violation effect for $\nu_\mu
\to\nu_e$ oscillation in $1.5 \le E \le 3$ ${\rm GeV}$ at $L = 290$km, and
is
at $(8-15)\%$ level for $\nu_\mu\to\nu_\tau$ oscillation in $6 \le E \le
10$ ${\rm GeV}$ at $L = 730$km for the active-sterile admixture range of
$0.15 \le D \le 0.8$, where $D \equiv |U_{s1}|^2+|U_{s2}|^2$.

Then, we have studied the relation between the active-sterile admixture
$(D)$
of neutrinos and the magnitude of pure CP violation effect. For the
close-to-active solar neutrino oscillations plus close-to-sterile
atmospheric
neutrino oscillations $(D \sim 0.2)$ \cite{Gonzalez-Garcia}, the pure CP
violation effect in $\Delta P(\nu_\mu\to\nu_\tau)$ could attain the
magnitude
as large as 0.10-0.25 in $6 \le E \le 15$ ${\rm GeV}$ at $L = 730$km for
$\Delta m^2_{\rm atm} = 2.5\times 10^{-3}{\rm eV}^2, \Delta m^2_{\rm solar}
=
5\times 10^{-5}{\rm eV}^2$ and $\Delta m^2_{\rm LSND} = 0.3{\rm eV}^2$. This
magnitude is prominently governed by the mixing angle product $c_{23}s_{23}$
in $\Delta P(\nu_\mu\to\nu_\tau)$, and $s_{23}$ determines the
active-sterile
admixture $D$. For near-pure-sterile solar neutrino oscillations plus
near-pure-active atmospheric neutrino oscillations $(D \sim 0.91-0.97)$
\cite{Gonzalez-Garcia}, the pure CP violation
effect in $\Delta P(\nu_\mu\to\nu_\tau)$ is very small, about 0.01, in
$6 \le E \le 15$ ${\rm GeV}$ at $L = 730$km and is comparable to the matter
effect. On the other hand, for $\nu_\mu\to\nu_e$ oscillation, the pure CP
violation effect is independent of the active-sterile admixture and is at
most
0.05 in $1.5 \le E \le 3$ ${\rm GeV}$ at $L = 290$km, which is almost the
same
in magnitude as in the three-neutrino model.

It may be interesting to measure the CP violation effect in $\nu_\mu
\to\nu_\tau$ oscillation for the baseline of $L=730$km by using the
conventional super-beams of $\nu_\mu$ and $\bar{\nu_\mu}$ in the energy
range
of $E = 6-15$ GeV \cite{MINOS} \cite{OPERA}. Also it might be intriguing to
measure the CP violation effect in $\nu_\mu\to\nu_e$ oscillation for
$L=250 -
300$km by using the conventional super-beams of $\nu_\mu$ and
$\bar{\nu_\mu}$
in $E = 0.1-3$ GeV \cite{JHF}. .
\vskip 0.7truecm

\centerline{\large\bf Appendix:  Oscillation probability}
\vskip 0.2truecm

Here we present the oscillation probability of Eq.(9) with Eq.(17) taken in
Eq.(16).
\begin{eqnarray}
&P&(\nu_\alpha\to\nu_\beta; L)    \nonumber  \\
&=& \delta_{\beta\alpha}[\> 1-4\sin^2\left( \frac{\Delta m^2_{31}L}{4E}
\right) \{ |U_{\alpha 3}|^2[\> 1-2\frac{a}{\Delta m^2_{31}}(|U_{e3}|^2
-\delta_{\alpha e})    \nonumber  \\
& & {} -2\frac{a'}{\Delta m^2_{31}}(|U_{s3}|^2-\delta_{\alpha s})\> ]
-2( \frac{a}{\Delta m^2_{31}}+\frac{a}{\Delta m^2_{43}} )
{\rm Re}(U_{\alpha 3}^*U_{\alpha 4}U_{e3}U_{e4}^*)
-2( \frac{a'}{\Delta m^2_{31}}   \nonumber  \\
& & {} +\frac{a'}{\Delta m^2_{43}} )
{\rm Re}(U_{\alpha 3}^*U_{\alpha 4}U_{s3}U_{s4}^*) \}
-2\sin\left( \frac{\Delta m^2_{31}L}{2E} \right)|U_{\alpha 3}|^2
( \frac{aL}{2E}|U_{e3}|^2+\frac{a'L}{2E}|U_{s3}|^2 )    \nonumber  \\
& & {} -4\sin^2\left( \frac{\Delta m^2_{41}L}{4E} \right) \{ |U_{\alpha
4}|^2
[\> 1-2\frac{a}{\Delta m^2_{41}}(|U_{e3}|^2-\delta_{\alpha e})
-2\frac{a'}{\Delta m^2_{41}}(|U_{s4}|^2-\delta_{\alpha s})\> ]   \nonumber
\\
& & {} -2( \frac{a}{\Delta m^2_{41}}-\frac{a}{\Delta m^2_{43}} )
{\rm Re}(U_{\alpha 3}^*U_{\alpha 4}U_{e3}U_{e4}^*)
-2( \frac{a'}{\Delta m^2_{41}}-\frac{a'}{\Delta m^2_{43}} )
{\rm Re}(U_{\alpha 3}^*U_{\alpha 4}U_{s3}U_{s4}^*) \}    \nonumber  \\
& & {} -2\sin\left( \frac{\Delta m^2_{41}L}{2E} \right) |U_{\alpha 4}|^2
( \frac{aL}{2E}|U_{e4}|^2+\frac{a'L}{2E}|U_{s4}|^2 )\> ]    \nonumber  \\
& & {} +4\sin^2\left( \frac{\Delta m^2_{31}L}{4E }\right) [\> |U_{\alpha
3}|^2
|U_{\beta 3}|^2 \{1-2\frac{a}{\Delta m^2_{31}}(2|U_{e3}|^2
-\delta_{\alpha e}-\delta_{\beta e})    \nonumber  \\
& & {} -2\frac{a'}{\Delta m^2_{31}}(2|U_{s3}|^2-\delta_{\alpha s}
-\delta_{\beta s}) \}-2( \frac{a}{\Delta m^2_{31}}+\frac{a}
{\Delta m^2_{43}} )\{ |U_{\alpha 3}|^2{\rm Re}(U_{\beta 3}^*
U_{\beta 4}U_{e3}U_{e4}^*)    \nonumber  \\
& & {} +|U_{\beta 3}|^2{\rm Re}(U_{\alpha 3}^*U_{\alpha 4}U_{e3}
U_{e4}^*) \}-2( \frac{a'}{\Delta m^2_{31}}+\frac{a'}{\Delta m^2_{43}} )
\{|U_{\alpha 3}|^2{\rm Re}(U_{\beta 3}^*U_{\beta 4}U_{s3}U_{s4}^*)
\nonumber  \\
& & {} +|U_{\beta 3}|^2{\rm Re}(U_{\alpha 3}^*U_{\alpha 4}U_{s3}
U_{s4}^*) \}\> ]+2\frac{\Delta m^2_{31}L}{2E}\sin\left(
\frac{\Delta m^2_{31}L}{2E} \right)
[ \frac{\Delta m^2_{21}}{\Delta m^2_{31}}{\rm Re}(U_{\beta 3}^*
U_{\beta 2}U_{\alpha 3}U_{\alpha 2}^*)   \nonumber  \\
& & {} +\frac{a}{\Delta m^2_{31}}\{|U_{e3}|^2\delta_{\alpha e}
\delta_{\beta e}+|U_{\alpha 3}|^2|U_{\beta 3}|^2(2|U_{e3}|^2
-\delta_{\alpha e}-\delta_{\beta e})+{\rm Re}(U_{\beta 3}^*U_{\beta 4}
U_{\alpha 3}U_{\alpha 4}^*)   \nonumber  \\
& & {} \times (2|U_{e4}|^2-\delta_{\alpha e}-\delta_{\beta e})+|U_{\alpha
3}|^2
{\rm Re}(U_{\beta 3}^*U_{\beta 4}U_{e3}U_{e4}^*)+|U_{\beta 3}|^2
{\rm Re}(U_{\alpha 3}^*U_{\alpha 4}U_{e3}U_{e4}^*)\}   \nonumber  \\
& & {} +\frac{a'}{\Delta m^2_{31}}\{|U_{s3}|^2\delta_{\alpha s}
\delta_{\beta s}+|U_{\alpha 3}|^2|U_{\beta 3}|^2(2|U_{s3}|^2-\delta_{\alpha
s}
-\delta_{\beta s})+{\rm Re}(U_{\beta 3}^*U_{\beta 4}U_{\alpha 3}
U_{\alpha 4}^*)   \nonumber  \\
& & {} \times (2|U_{s4}|^2-\delta_{\alpha s}-\delta_{\beta s})+|U_{\alpha
3}|^2
{\rm Re}(U_{\beta 3}^*U_{\beta 4}U_{s3}U_{s4}^*)+|U_{\beta 3}|^2
{\rm Re}(U_{\alpha 3}^*U_{\alpha 4}U_{s3}U_{s4}^*)\} ]
\nonumber  \\
& & {} -4\frac{\Delta m^2_{31}L}{2E}\sin^2\left( \frac{\Delta m^2_{31}L}
{4E} \right) [\> \frac{\Delta m^2_{21}}{\Delta m^2_{31}}{\rm Im}
(U_{\beta 3}^*U_{\beta 2}U_{\alpha 3}U_{\alpha 2}^*)+\frac{a}
{\Delta m^2_{31}}\{{\rm Im}(U_{\beta 3}^*U_{\beta 4}U_{\alpha 3}
U_{\alpha 4}^*)   \nonumber  \\
& & {} \times (2|U_{e4}|^2-\delta_{\alpha e}-\delta_{\beta e})+|U_{\alpha
3}|^2
{\rm Im}(U_{\beta 3}^*U_{\beta 4}U_{e3}U_{e4}^*)-|U_{\beta 3}|^2
{\rm Im}(U_{\alpha 3}^*U_{\alpha 4}U_{e3}U_{e4}^*)\}   \nonumber  \\
& & {} +\frac{a'}{\Delta m^2_{31}}\{ {\rm Im}(U_{\beta 3}^*U_{\beta 4}
U_{\alpha 3}U_{\alpha 4}^*)(2|U_{s4}|^2-\delta_{\alpha s}
-\delta_{\beta s})+|U_{\alpha 3}|^2{\rm Im}(U_{\beta 3}^*U_{\beta 4}
U_{s3}U_{s4}^*)   \nonumber  \\
& & {} -|U_{\beta 3}|^2{\rm Im}(U_{\alpha 3}^*U_{\alpha 4}U_{s3}
U_{s4}^*)\}\> ]   \nonumber  \\
& & {} +4\sin^2\left( \frac{\Delta m^2_{41}L}{4E} \right) [\>
|U_{\alpha 4}|^2|U_{\beta 4}|^2 \{1-2\frac{a}{\Delta m^2_{41}}(2|U_{e4}|^2
-\delta_{\alpha e}-\delta_{\beta e})    \nonumber  \\
& & {} -2\frac{a'}{\Delta m^2_{41}}(2|U_{s4}|^2-\delta_{\alpha s}
-\delta_{\beta s}) \}-2( \frac{a}{\Delta m^2_{41}}-\frac{a}
{\Delta m^2_{43}} ) \{|U_{\alpha 4}|^2{\rm Re}(U_{\beta 3}^*
U_{\beta 4}U_{e3}U_{e4}^*)    \nonumber  \\
& & {} +|U_{\beta 4}|^2{\rm Re}(U_{\alpha 3}^*U_{\alpha 4}U_{e3}
U_{e4}^*)\}-2( \frac{a'}{\Delta m^2_{41}}-\frac{a'}{\Delta m^2_{43}} )
\{|U_{\alpha 4}|^2{\rm Re}(U_{\beta 3}^*U_{\beta 4}U_{s3}U_{s4}^*)
\nonumber  \\
& & {} +|U_{\beta 4}|^2{\rm Re}(U_{\alpha 3}^*U_{\alpha 4}U_{s3}
U_{s4}^*)\}]+2\frac{\Delta m^2_{31}L}{2E}\sin\left(
\frac{\Delta m^2_{41}L}{2E} \right) [\> \frac{\Delta m^2_{21}}
{\Delta m^2_{31}}{\rm Re}(U_{\beta 4}^*
U_{\beta 2}U_{\alpha 4}U_{\alpha 2}^*)   \nonumber  \\
& & {} +\frac{a}{\Delta m^2_{31}}\{|U_{e4}|^2\delta_{\alpha e}
\delta_{\beta e}+|U_{\alpha 4}|^2|U_{\beta 4}|^2(2|U_{e4}|^2-\delta_{\alpha
e}
-\delta_{\beta e})+{\rm Re}(U_{\beta 3}^*U_{\beta 4}U_{\alpha 3}
U_{\alpha 4}^*)   \nonumber  \\
& & {} \times (2|U_{e3}|^2-\delta_{\alpha e}-\delta_{\beta e})+|U_{\alpha
4}|^2
{\rm Re}(U_{\beta 3}^*U_{\beta 4}U_{e3}U_{e4}^*)+|U_{\beta 4}|^2
{\rm Re}(U_{\alpha 3}^*U_{\alpha 4}U_{e3}U_{e4}^*)\}   \nonumber  \\
& & {} +\frac{a'}{\Delta m^2_{31}} \{|U_{s4}|^2\delta_{\alpha s}
\delta_{\beta s}+|U_{\alpha 4}|^2|U_{\beta 4}|^2(2|U_{s4}|^2-\delta_{\alpha
s}
-\delta_{\beta s})+{\rm Re}(U_{\beta 3}^*U_{\beta 4}U_{\alpha 3}
U_{\alpha 4}^*)   \nonumber  \\
& & {} \times (2|U_{s3}|^2-\delta_{\alpha s}-\delta_{\beta s})+|U_{\alpha
4}|^2
{\rm Re}(U_{\beta 3}^*U_{\beta 4}U_{s3}U_{s4}^*)+|U_{\beta 4}|^2
{\rm Re}(U_{\alpha 3}^*U_{\alpha 4}U_{s3}U_{s4}^*)\}\> ]
\nonumber  \\
& & {} -4\frac{\Delta m^2_{31}L}{2E}\sin^2\left( \frac{\Delta m^2_{41}L}
{4E} \right) [\> \frac{\Delta m^2_{21}}{\Delta m^2_{31}}
{\rm Im}(U_{\beta 4}^*U_{\beta 2}U_{\alpha 4}U_{\alpha 2}^*)
+\frac{a}{\Delta m^2_{31}}\{{\rm Im}(U_{\beta 4}^*U_{\beta 3}
U_{\alpha 4}U_{\alpha 3}^*)   \nonumber  \\
& & {} \times (2|U_{e3}|^2-\delta_{\alpha e}-\delta_{\beta e})+|U_{\alpha
4}|^2
{\rm Im}(U_{\beta 4}^*U_{\beta 3}U_{e4}U_{e3}^*)-|U_{\beta 4}|^2
{\rm Im}(U_{\alpha 4}^*U_{\alpha 3}U_{e4}U_{e3}^*) \}   \nonumber  \\
& & {} +\frac{a'}{\Delta m^2_{31}} \{{\rm Im}(U_{\beta 4}^*U_{\beta 3}
U_{\alpha 4}U_{\alpha 3}^*)(2|U_{s3}|^2-\delta_{\alpha s}-\delta_{\beta s})
+|U_{\alpha 4}|^2{\rm Im}(U_{\beta 4}^*U_{\beta 3}U_{s4}U_{s3}^*)
\nonumber  \\
& & {} -|U_{\beta 4}|^2{\rm Im}(U_{\alpha 4}^*U_{\alpha 3}U_{s4}
U_{s3}^*)\}\> ]   \nonumber  \\
& & {} +8\sin\left( \frac{\Delta m^2_{31}L}{4E} \right)\sin\left(
\frac{\Delta m^2_{41}L}{4E} \right)\cos\left( \frac{\Delta m^2_{43}L}{4E}
\right)  [\> {\rm Re}(U_{\beta 3}^*U_{\beta 4}
U_{\alpha 3}U_{\alpha 4}^*)    \nonumber  \\
& & {} \times \{ 1-\frac{a}{\Delta m^2_{31}}(2|U_{e3}|^2-\delta_{\alpha e}
-\delta_{\beta e})-\frac{a}{\Delta m^2_{41}}(2|U_{e4}|^2-\delta_{\alpha e}
-\delta_{\beta e})    \nonumber  \\
& & {} -\frac{a'}{\Delta m^2_{31}}(2|U_{s3}|^2 -\delta_{\alpha s}
-\delta_{\beta s})-\frac{a'}{\Delta m^2_{41}}(2|U_{s4}|^2 -\delta_{\alpha s}
-\delta_{\beta s}) \}    \nonumber  \\
& & {} +{\rm Im}(U_{\beta 3}^*U_{\beta 4}U_{\alpha 3}U_{\alpha 4}^*)
( \frac{aL}{2E}|U_{e4}|^2+\frac{a'L}{2E}|U_{s4}|^2-\frac{aL}{2E}
|U_{e3}|^2-\frac{a'L}{2E}|U_{s3}|^2 )    \nonumber  \\
& & {} -( \frac{a}{\Delta m^2_{41}}-\frac{a}{\Delta m^2_{43}} )
\{|U_{\alpha 3}|^2{\rm Re}(U_{\beta 3}^*U_{\beta 4}U_{e3}U_{e4}^*)
+|U_{\beta 3}|^2{\rm Re}(U_{\alpha 3}^*U_{\alpha 4}U_{e3}U_{e4}^*)\}
\nonumber  \\
& & {} -( \frac{a'}{\Delta m^2_{41}}-\frac{a'}{\Delta m^2_{43}} )
\{|U_{\alpha 3}|^2{\rm Re}(U_{\beta 3}^*U_{\beta 4}U_{s3}U_{s4}^*)
+|U_{\beta 3}|^2{\rm Re}(U_{\alpha 3}^*U_{\alpha 4}U_{s3}U_{s4}^*)\}
\nonumber  \\
& & {} -( \frac{a}{\Delta m^2_{31}}+\frac{a}{\Delta m^2_{43}} )
\{|U_{\alpha 4}|^2{\rm Re}(U_{\beta 3}^*U_{\beta 4}U_{e3}U_{e4}^*)
+|U_{\beta 4}|^2{\rm Re}(U_{\alpha 3}^*U_{\alpha 4}U_{e3}U_{e4}^*)\}
\nonumber  \\
& & {} -( \frac{a'}{\Delta m^2_{31}}+\frac{a'}{\Delta m^2_{43}} )
\{|U_{\alpha 4}|^2{\rm Re}(U_{\beta 3}^*U_{\beta 4}U_{s3}U_{s4}^*)
+|U_{\beta 4}|^2{\rm Re}(U_{\alpha 3}^*U_{\alpha 4}U_{s3}
U_{s4}^*)\}\> ]     \nonumber  \\
& & {} +8\sin\left( \frac{\Delta m^2_{31}L}{4E} \right)\sin\left(
\frac{\Delta m^2_{41}L}{4E} \right)\sin\left( \frac{\Delta m^2_{43}L}{4E}
\right) [\> {\rm Im}(U_{\beta 3}^*U_{\beta 4}U_{\alpha 3}U_{\alpha 4}^*)
\nonumber  \\
& & {} \times \{1-\frac{a}{\Delta m^2_{31}}(2|U_{e3}|^2-\delta_{\alpha e}
-\delta_{\beta e})-\frac{a}{\Delta m^2_{41}}(2|U_{e4}|^2-\delta_{\alpha e}
-\delta_{\beta e})     \nonumber  \\
& & {} -\frac{a'}{\Delta m^2_{31}}(2|U_{s3}|^2 -\delta_{\alpha s}
-\delta_{\beta s})-\frac{a'}{\Delta m^2_{41}}(2|U_{s4}|^2 -\delta_{\alpha s}
-\delta_{\beta s}) \}    \nonumber  \\
& & {} -{\rm Re}(U_{\beta 3}^*U_{\beta 4}U_{\alpha 3}U_{\alpha 4}^*)
( \frac{aL}{2E}|U_{e4}|^2+\frac{a'L}{2E}|U_{s4}|^2-\frac{aL}{2E}
|U_{e3}|^2-\frac{a'L}{2E}|U_{s3}|^2 )    \nonumber  \\
& & {} -( \frac{a}{\Delta m^2_{41}}-\frac{a}{\Delta m^2_{43}} )
\{|U_{\alpha 3}|^2{\rm Im}(U_{\beta 3}^*U_{\beta 4}U_{e3}U_{e4}^*)
-|U_{\beta 3}|^2{\rm Im}(U_{\alpha 3}^*U_{\alpha 4}U_{e3}U_{e4}^*)\}
\nonumber  \\
& & {} -( \frac{a'}{\Delta m^2_{41}}-\frac{a'}{\Delta m^2_{43}} )
\{|U_{\alpha 3}|^2{\rm Im}(U_{\beta 3}^*U_{\beta 4}U_{s3}U_{s4}^*)
-|U_{\beta 3}|^2{\rm Im}(U_{\alpha 3}^*U_{\alpha 4}U_{s3}U_{s4}^*)\}
\nonumber  \\
& & {} -( \frac{a}{\Delta m^2_{31}}+\frac{a}{\Delta m^2_{43}} )
\{|U_{\alpha 4}|^2{\rm Im}(U_{\beta 3}^*U_{\beta 4}U_{e3}U_{e4}^*)
-|U_{\beta 4}|^2{\rm Im}(U_{\alpha 3}^*U_{\alpha 4}U_{e3}U_{e4}^*)\}
\nonumber  \\
& & {} -( \frac{a'}{\Delta m^2_{31}}+\frac{a'}{\Delta m^2_{43}} )
\{|U_{\alpha 4}|^2{\rm Im}(U_{\beta 3}^*U_{\beta 4}U_{s3}U_{s4}^*)
-|U_{\beta 4}|^2{\rm Im}(U_{\alpha 3}^*U_{\alpha 4}U_{s3}U_{s4}^*)\}
\> ] .     \nonumber
\end{eqnarray}

\newpage
\begin{figure}
\begin{center}
\leavevmode
\epsfxsize=14cm
\epsfbox{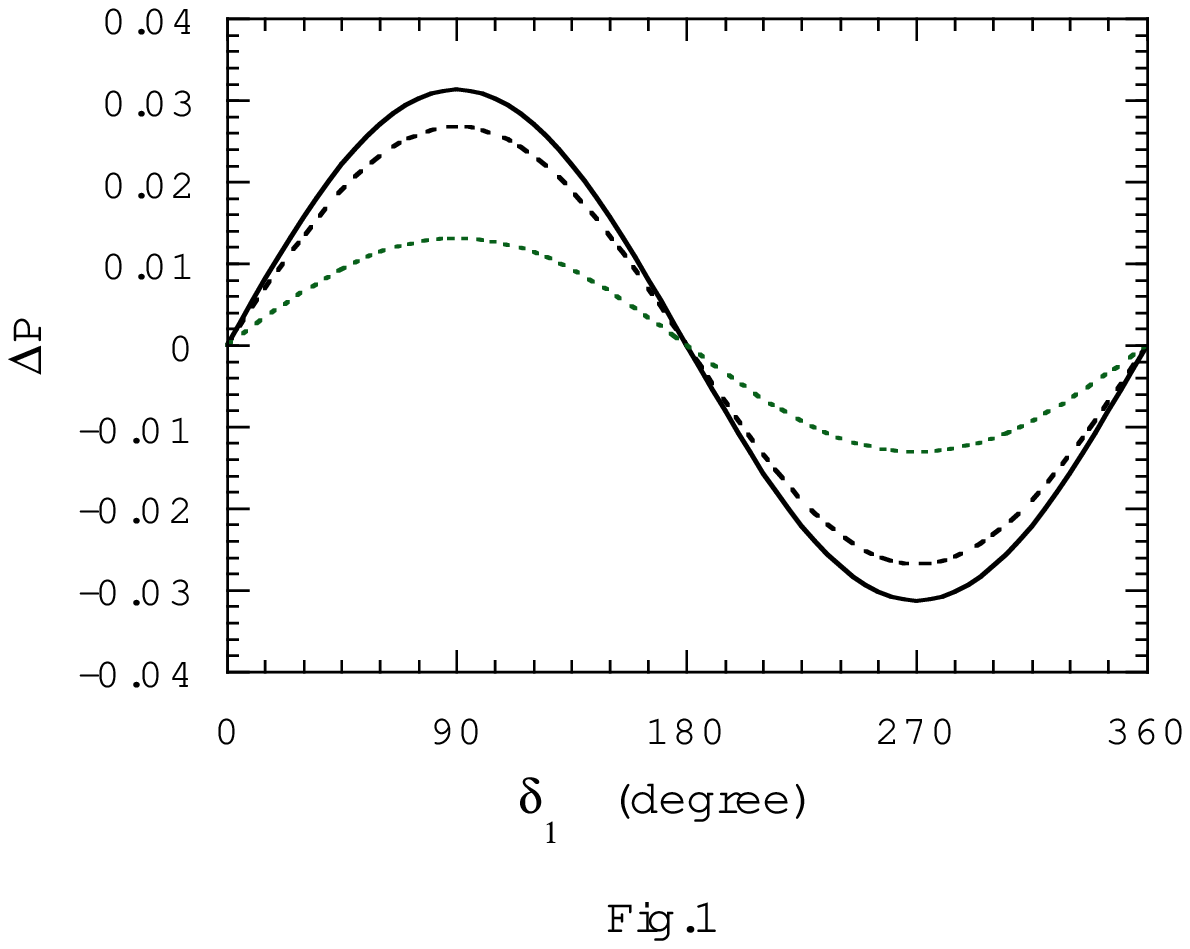}
\end{center}
\caption{The pure CP violation effect in  $\nu_{\mu}\to\nu_e$ oscillation
 with respect to the phase $\delta_1$ of the mixing matrix in the
long-baseline experiment with the baseline $L=290$ km and the neutrino energy
$E=1.2$ GeV for the typical three parameter sets; $(s_{02}=0.12, s_{03}=0.06,
s_{12}=0.93, s_{13}=0.71)$(solid line), $(s_{02}=0.12, s_{03}=0.05, s_{12}
=0.97, s_{13}=0.71)$(dashed line), and $(s_{02}=0.15, s_{03}=0.02,
s_{12}=0.95,
s_{13}=0.71)$(dotted line), and commonly taken as $s_{01}=s_{23}=1/\sqrt{2},
\delta_{01}=\delta_{02}=\delta_{03}=\delta_{12}=0$, and $\delta_2=\pi /2$.
The mass-squared differences of neutrinos are fixed as $\Delta m^2_{\rm atm}=
2.5\times 10^{-3}{\rm eV^2}, \Delta m^2_{\rm LSND}=0.3{\rm eV^2}$, and
$\Delta m^2_{\rm solar}=5\times 10^{-5}{\rm eV^2}$.}
\end{figure}
\vskip 0.5truecm

\newpage
\begin{figure}
\begin{center}
\leavevmode
\epsfxsize=14cm
\epsfbox{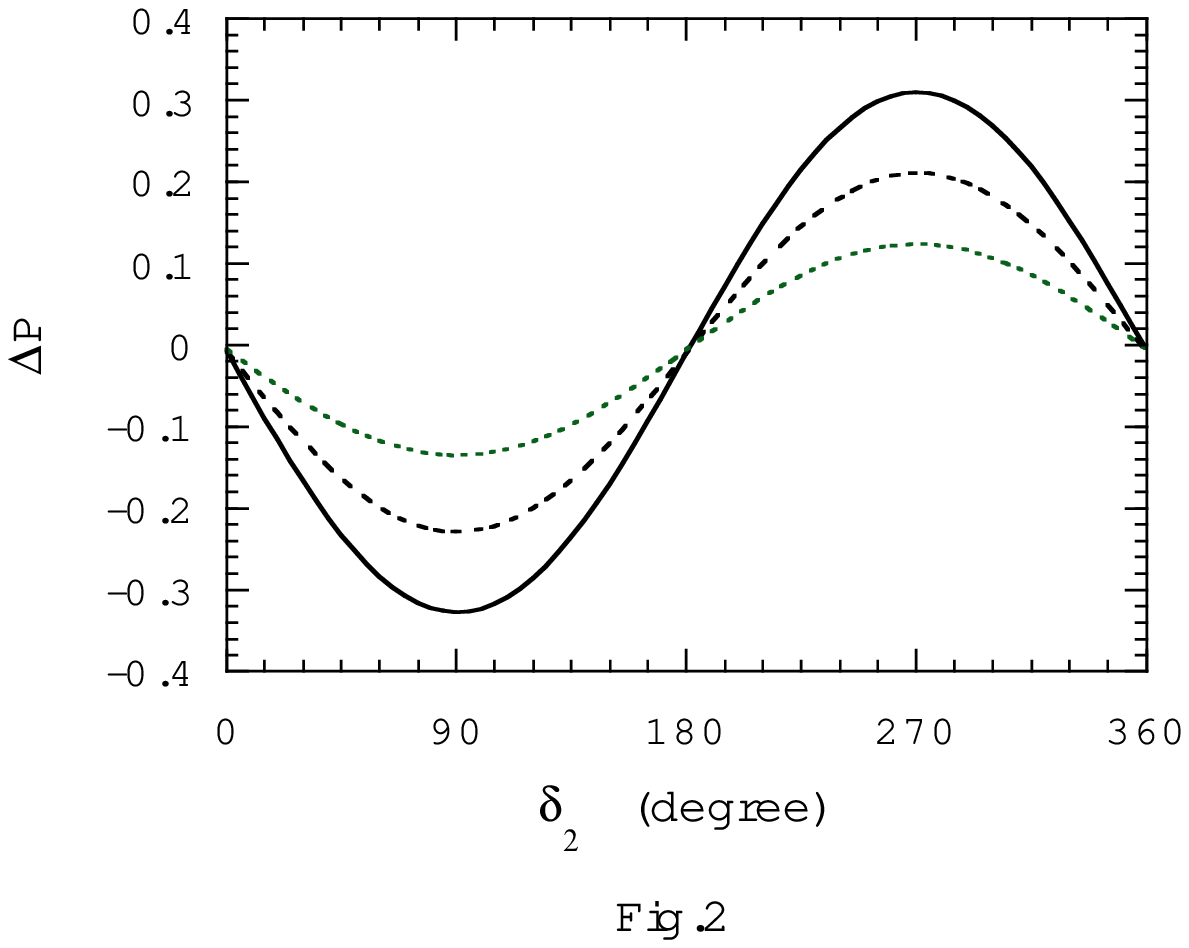}
\end{center}
\caption{The pure CP violation effect in  $\nu_{\mu}\to\nu_\tau$
oscillation with respect to the phase $\delta_2$ of the mixing matrix in the
long-baseline experiment with the baseline $L=730$ km and the neutrino energy
$E=6.1$ GeV for the typical three parameter sets; $(s_{02}=0.12, s_{03}=0.06,
s_{12}=0.93, s_{13}=0.71)$(solid line), $(s_{02}=0.12, s_{03}=0.06, s_{12}
=0.97, s_{13}=0.71)$(dashed line), and $(s_{02}=0.15, s_{03}=0.03,
s_{12}=0.99,
s_{13}=0.71)$(dotted line), and commonly taken as $s_{01}=s_{23}=1/\sqrt{2},
\delta_{01}=\delta_{02}=\delta_{03}=\delta_{12}=0$, and $\delta_1=\pi /2$.
The mass-squared differences of neutrinos are the same as in Fig.1.}
\end{figure}
\vskip 0.5truecm

\newpage
\begin{figure}
\begin{center}
\leavevmode
\epsfxsize=14cm
\epsfbox{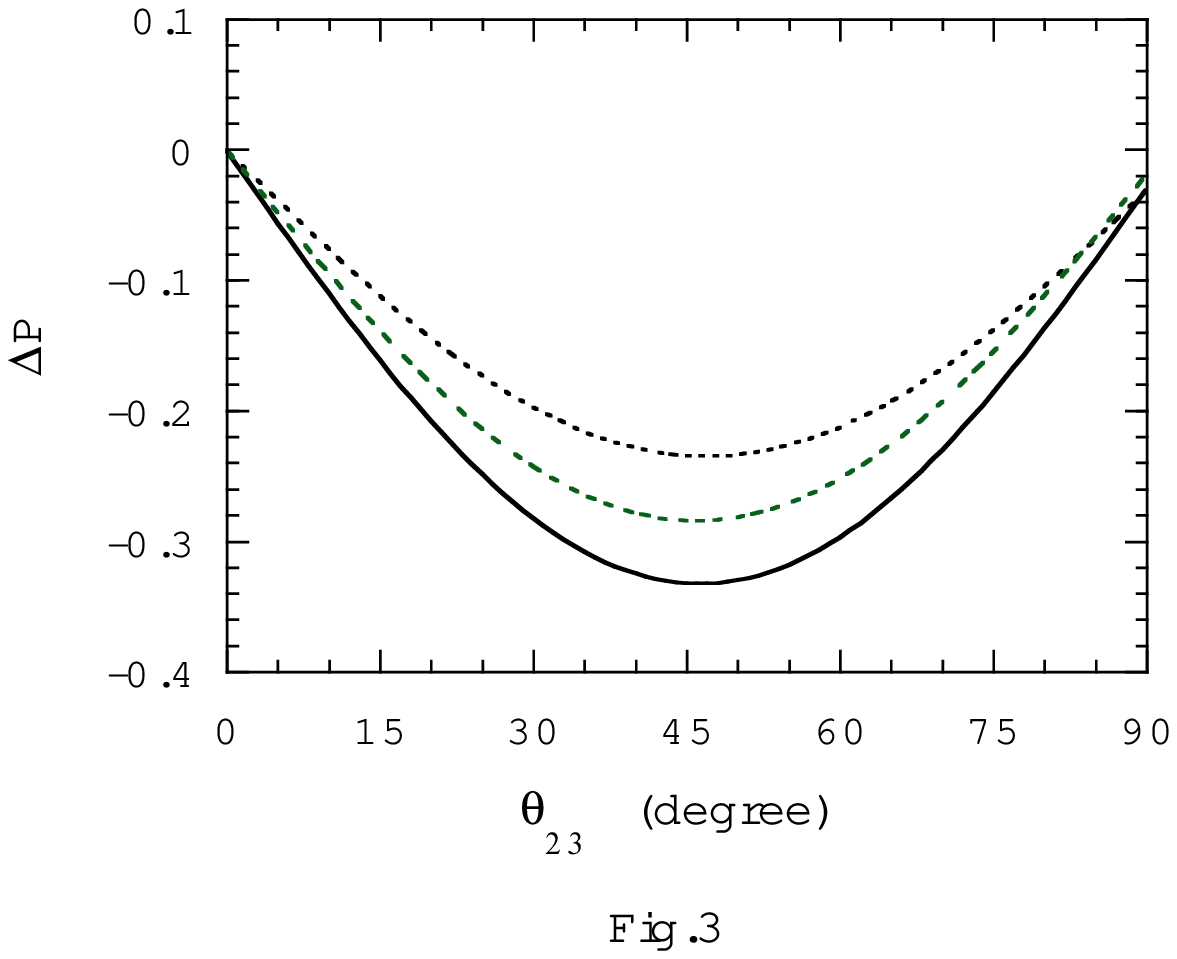}
\end{center}
\caption{The pure CP violation effect in  $\nu_{\mu}\to\nu_\tau$
oscillation with respect to the mixing angle $\theta_{23}$ of the mixing
matrix at $L=730$ km and $E=6.1$ GeV for the typical three parameter sets;
$(s_{02}=s_{03}=0.11, s_{12}=0.93, s_{13}=0.71)$(solid line), $(s_{02}=0.15,
s_{03}=0.05, s_{12}=0.95, s_{13}=0.71)$(dashed line), and $(s_{02}=s_{03}
=0.11, s_{12}=0.97, s_{13}=0.71)$(dotted line), and commonly taken as
$s_{01}
=1/\sqrt{2}, \delta_{01}=\delta_{02}=\delta_{03}=\delta_{12}=0$, and
$\delta_1=\delta_2=\pi /2$.}
\end{figure}
\vskip 0.5truecm

\newpage
\begin{figure}
\begin{center}
\leavevmode
\epsfxsize=14cm
\epsfbox{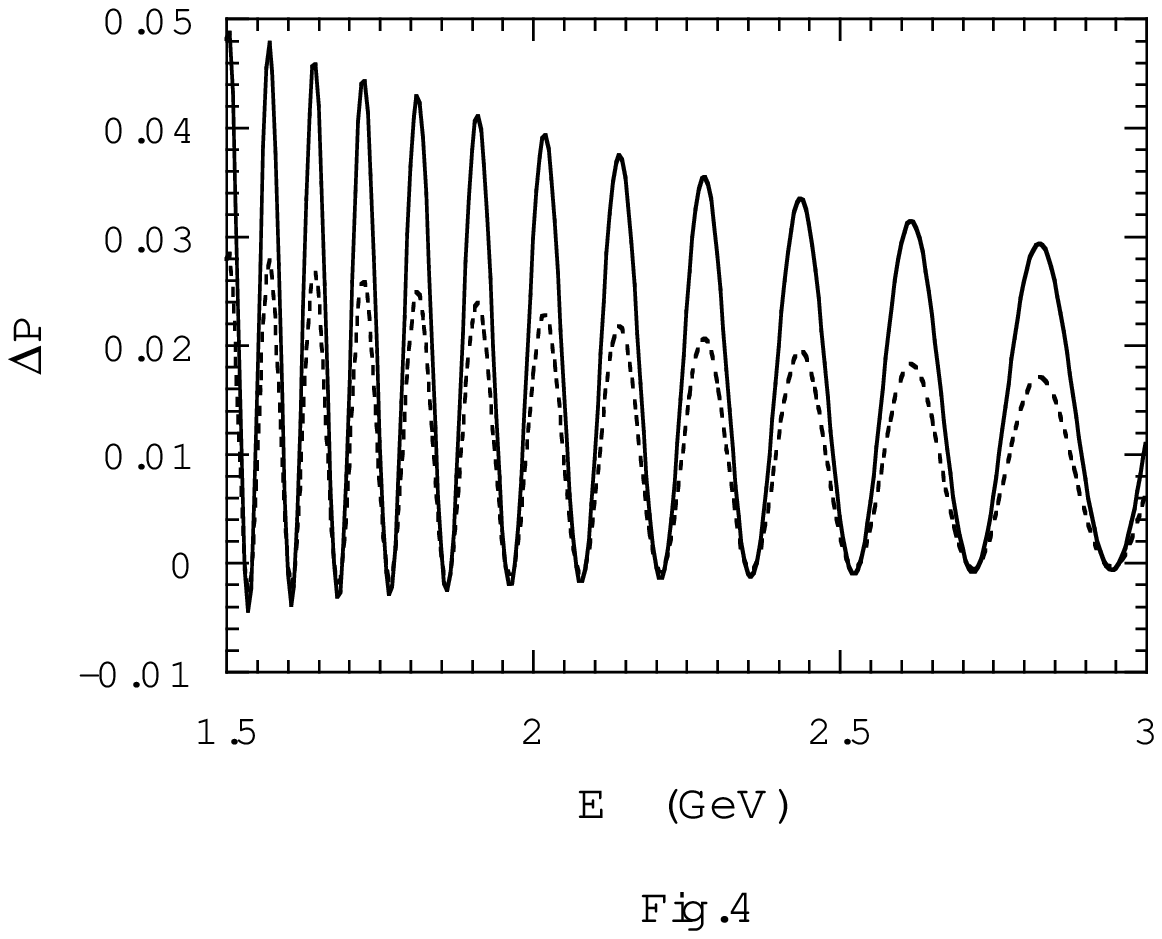}
\end{center}
\caption{The pure CP violation effect in $\nu_{\mu}\to\nu_e$ oscillation
 with respect to the neutrino energy $E$ (= 1.5-3 GeV) at $L=290$ km for the
typical two parameter sets; $(s_{02}=s_{03}=0.11, s_{12}=0.97, s_{13}=0.73)$
(solid line) and $(s_{02}=0.12, s_{03}=0.06, s_{12}=0.93, s_{13}=0.71)$(dashed
line), and commonly taken as $s_{01}=1/\sqrt{2}, s_{23}=0.4, \delta_{01}
=\delta_{02}=\delta_{03}=\delta_{12}=0$, and $\delta_1=\delta_2=\pi /2$.}
\end{figure}
\vskip 0.5truecm

\newpage
\begin{figure}
\begin{center}
\leavevmode
\epsfxsize=14cm
\epsfbox{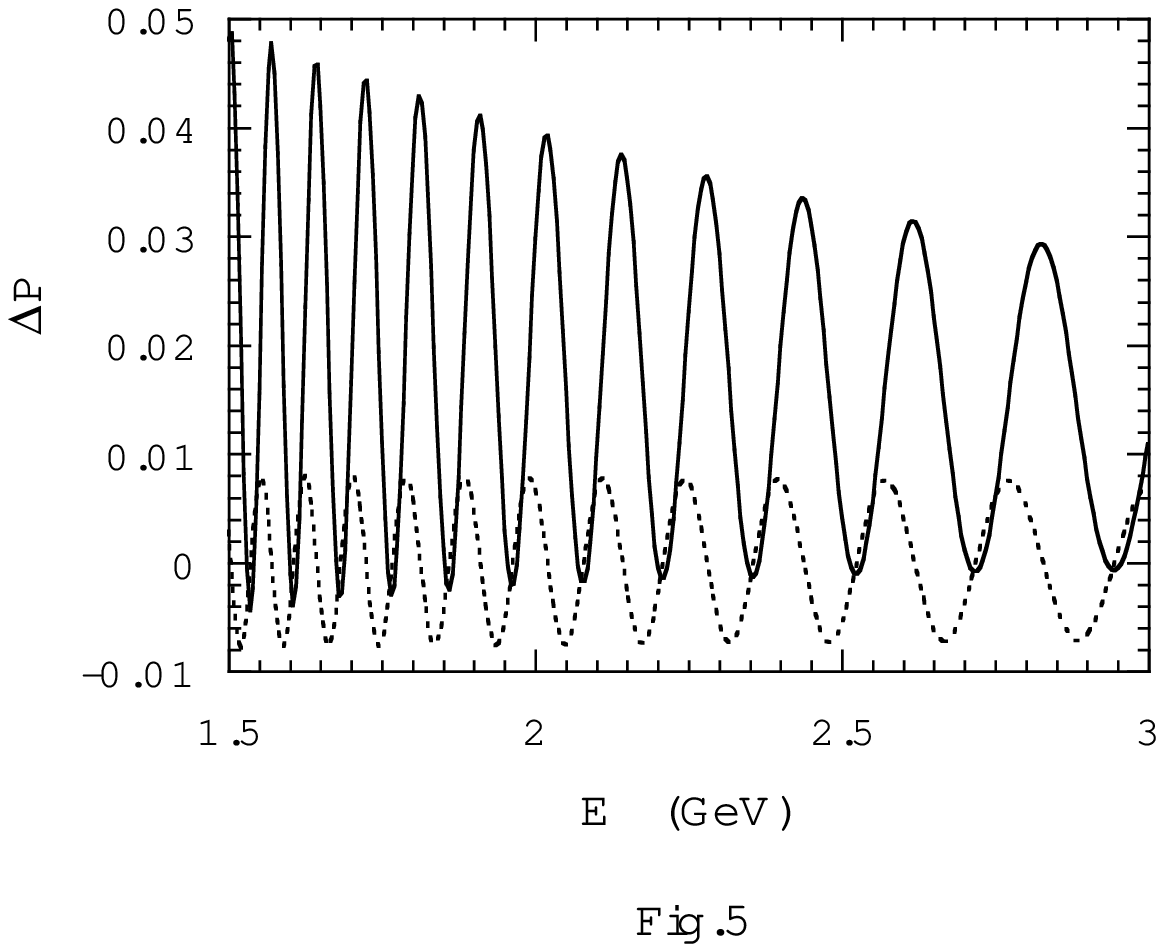}
\end{center}
\caption{The pure CP violation effect (solid line) and the matter effect
(dotted line) in  $\nu_{\mu}\to\nu_e$ oscillation with respect to the neutrino
energy $E$ at $L=290$ km for the parameter set, $(s_{02}=s_{03}=0.11, s_{12}
=0.97, s_{13}=0.73)$ and the other angles and phases are the same as inFig.4.}
\end{figure}
\vskip 0.5truecm

\newpage
\begin{figure}
\begin{center}
\leavevmode
\epsfxsize=14cm
\epsfbox{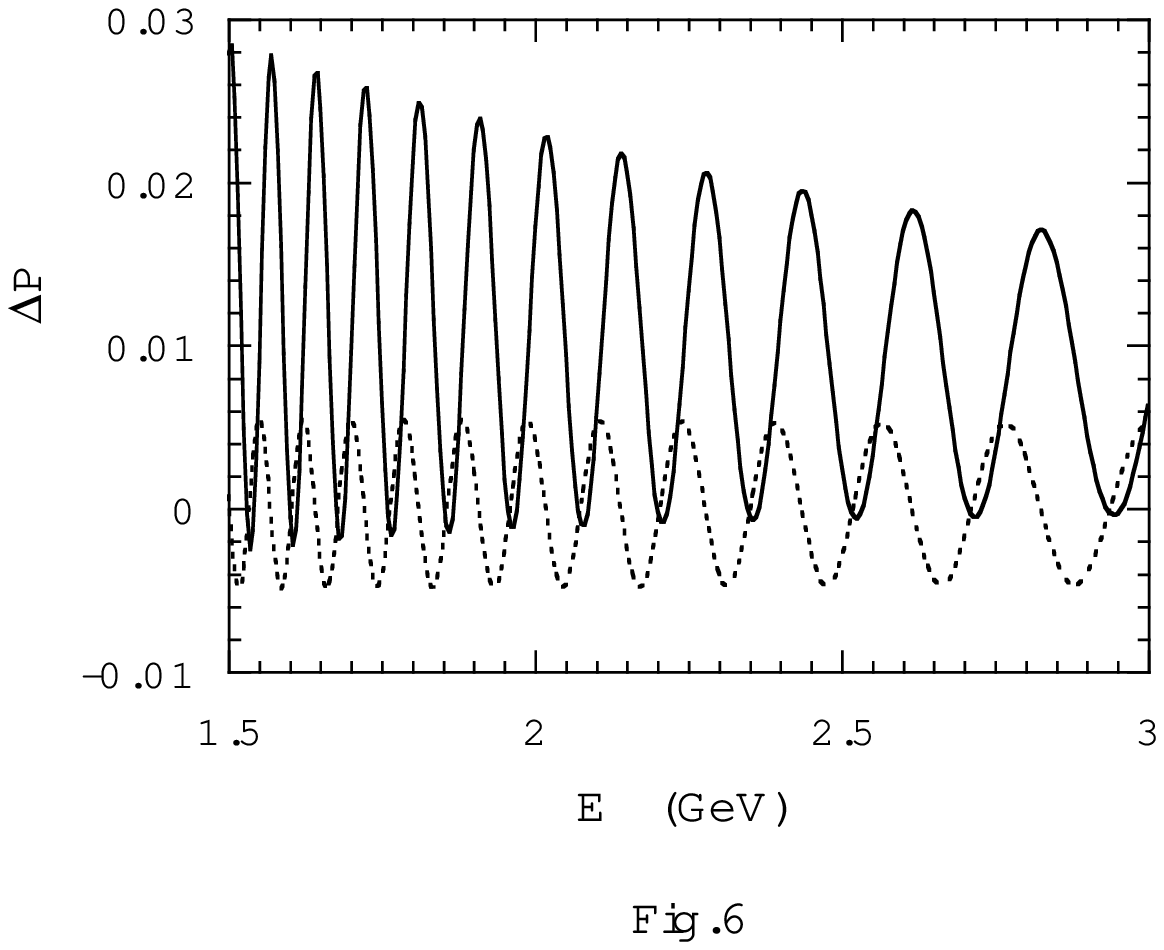}
\end{center}
\caption{The pure CP violation effect (solid line) and the matter effect
(dotted line) in  $\nu_{\mu}\to\nu_e$ oscillation with respect to the neutrino
energy $E$ at $L=290$ km for the parameter set, $(s_{02}=0.12, s_{03}=0.06,
s_{12}=0.93, s_{13}=0.71)$ and the other angles and phases are the same as
in Fig.4.}
\end{figure}
\vskip 0.5truecm

\newpage
\begin{figure}
\begin{center}
\leavevmode
\epsfxsize=14cm
\epsfbox{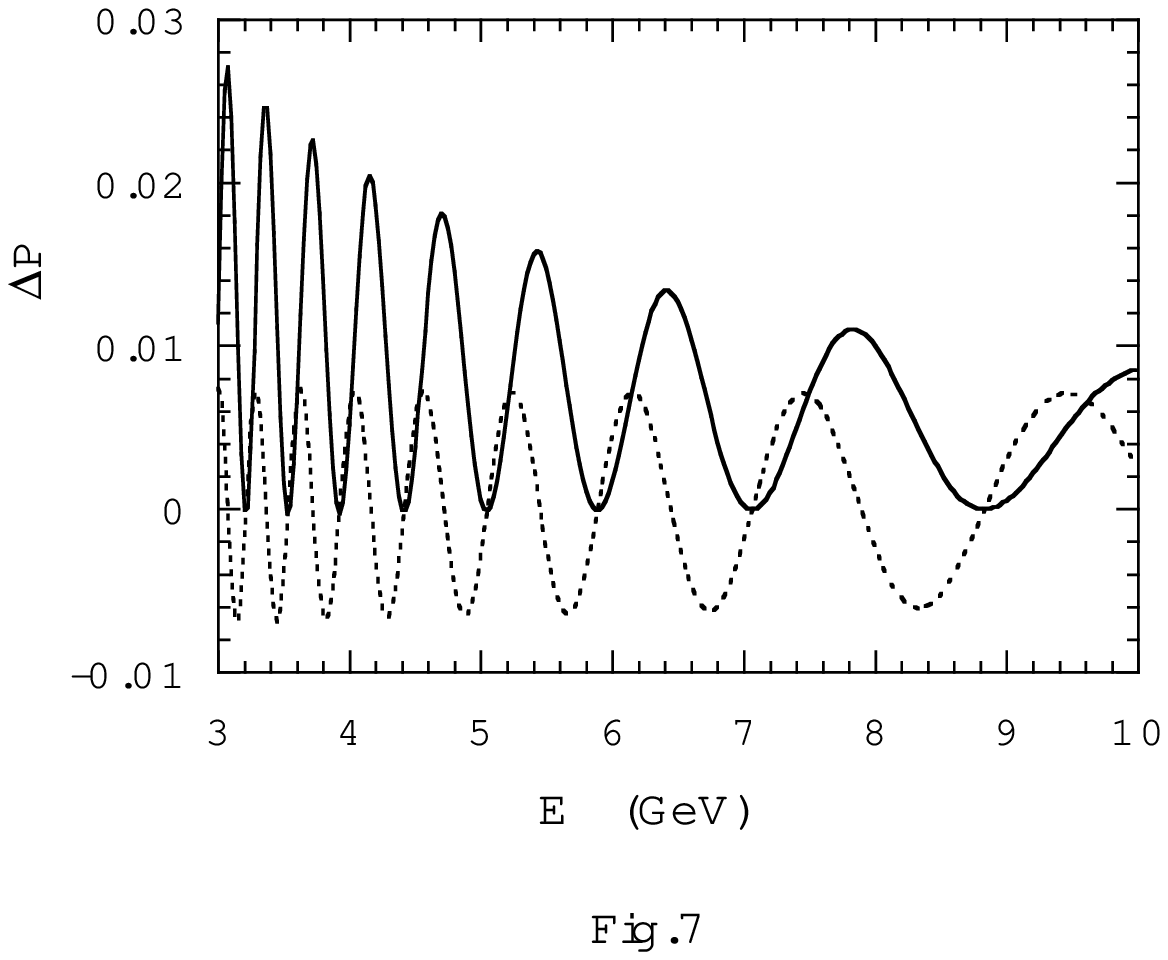}
\end{center}
\caption{The pure CP violation effect (solid line) and the matter effect
(dotted line) in  $\nu_{\mu}\to\nu_e$ oscillation in the energy range $E=3-10$
GeV at $L=290$ km for the parameter set, $(s_{02}=s_{03}=0.11, s_{12}=0.97,
s_{13}=0.73)$ and the other angles and phases are the same as in Fig.4.}
\end{figure}
\vskip 0.5truecm

\newpage
\begin{figure}
\begin{center}
\leavevmode
\epsfxsize=14cm
\epsfbox{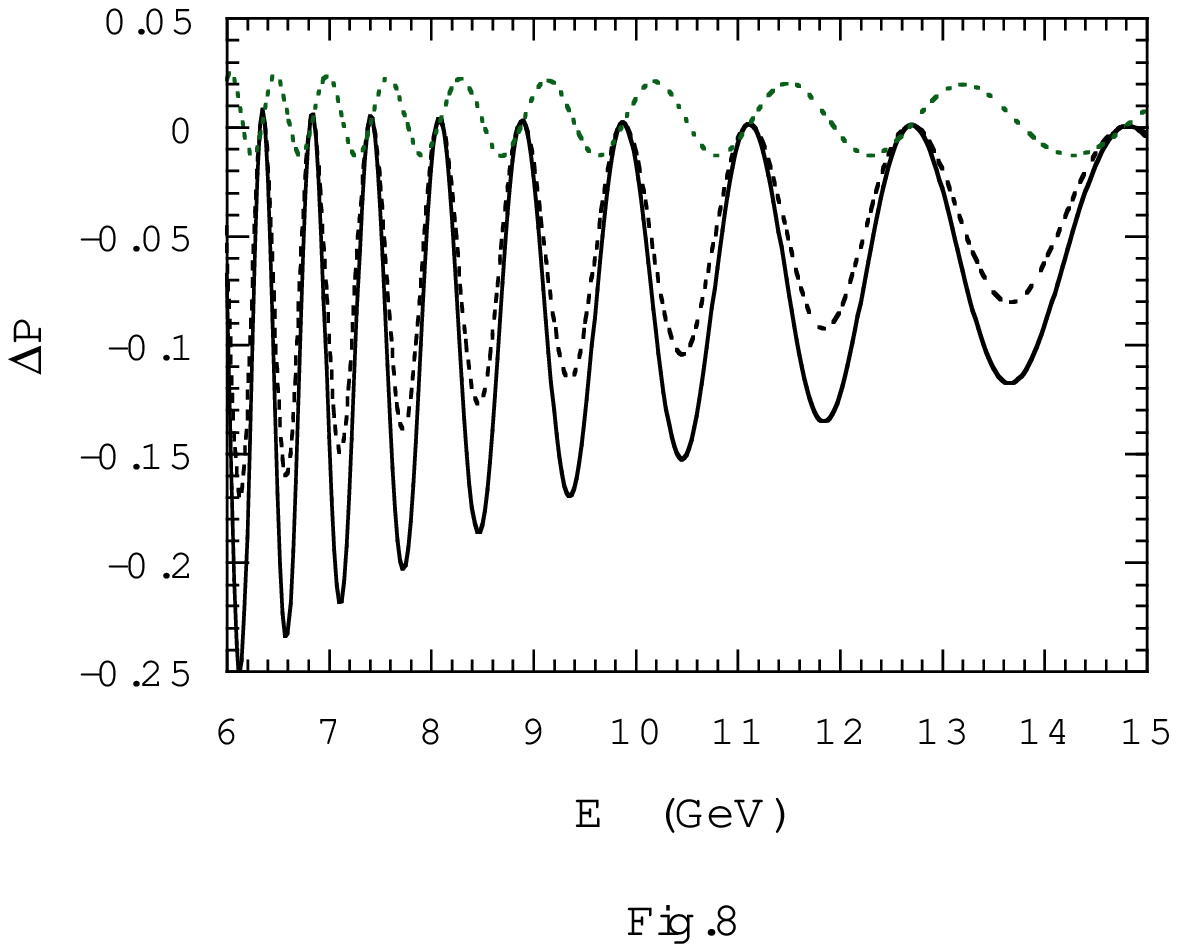}
\end{center}
\caption{The pure CP violation effect in $\nu_{\mu}\to\nu_\tau$
oscillation in the energy range of $E=6-15$ GeV at $L=730$ km for the typical
two parameter sets; $(s_{02}=0.12, s_{03}=0.06, s_{12}=0.93, s_{13}=0.71,
s_{23}=0.40)$(solid line) and $(s_{02}=0.12, s_{03}=0.06, s_{12}=0.97,
s_{13}=0.72, s_{23}=0.40)$(dashed line) for the active-sterile admixture
$D \sim 0.2$, and the matter effect (dotted line) for the first parameter set
of the above, and commonly taken as $s_{01}=1/\sqrt{2}, \delta_{01}
=\delta_{02}=\delta_{03}=\delta_{12}=0$, and $\delta_1=\delta_2=\pi /2$.}
\end{figure}
\vskip 0.5truecm

\newpage
\begin{figure}
\begin{center}
\leavevmode
\epsfxsize=14cm
\epsfbox{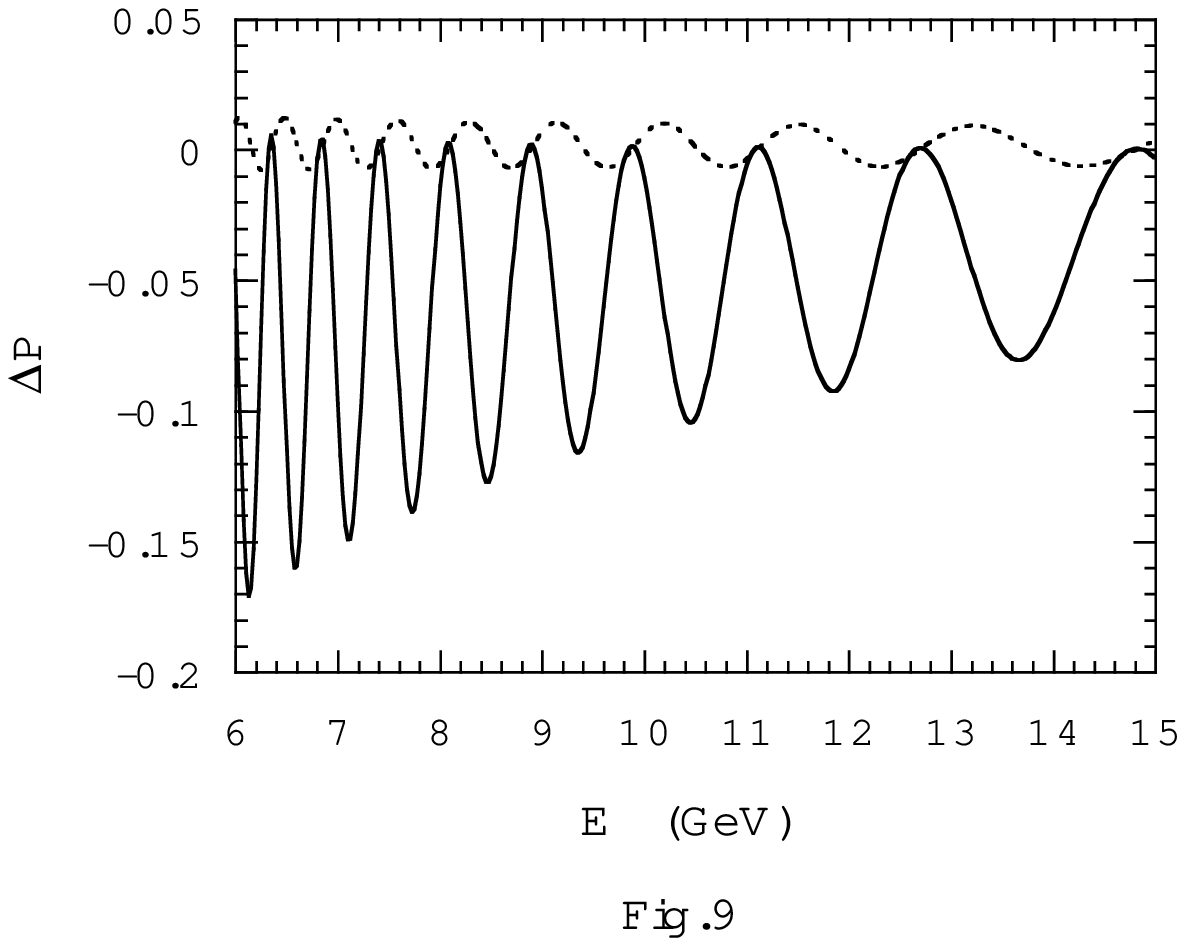}
\end{center}
\caption{
The pure CP violation effect (solid line) and the matter effect
(dotted line) in $\nu_{\mu}\to\nu_\tau$ oscillation at $L=730$ km for the
parameter set, $(s_{02}=0.12, s_{03}=0.06, s_{12}=0.97, s_{13}=0.72,
s_{23}=0.40)$ and the other angles and phases are the same as in Fig.8.}
\end{figure}
\vskip 0.5truecm

\newpage
\begin{figure}
\begin{center}
\leavevmode
\epsfxsize=14cm
\epsfbox{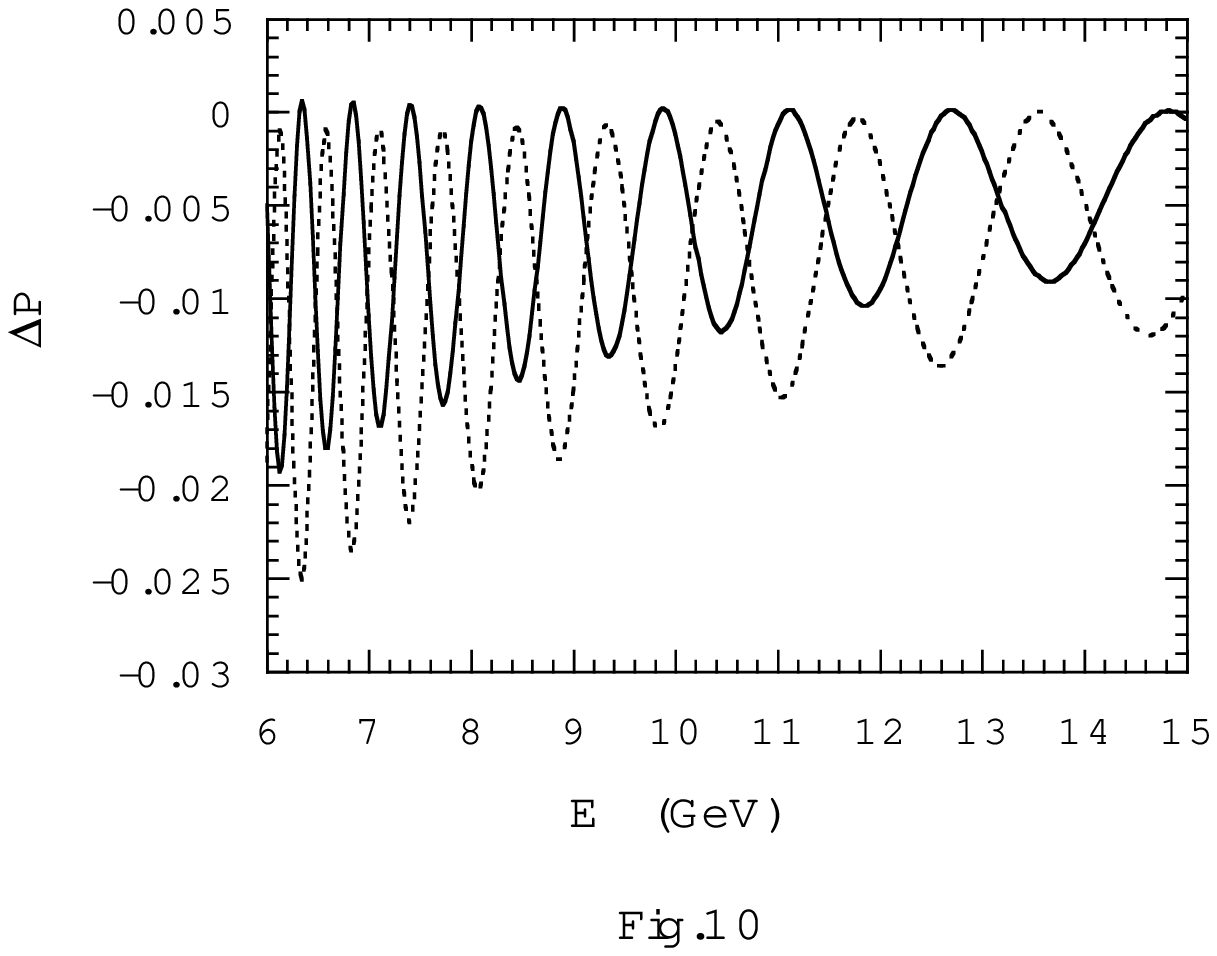}
\end{center}
\caption{The pure CP violation effect (solid line) and the matter effect
(dotted line) for the active-sterile admixture $D \sim 0.9$ in  $\nu_{\mu}
\to\nu_\tau$ oscillation at $L=730$ km for the parameter set, $(s_{02}=0.12,
s_{03}=0.06, s_{12}=0.95, s_{13}=0.71, s_{23}=1.0)$ and taken as $s_{01}
=1/\sqrt{2}, \delta_{01}=\delta_{02}=\delta_{03}=\delta_{12}=0$, and
$\delta_1=\delta_2=\pi /2$.}
\end{figure}
\vskip 0.5truecm

\newpage
\begin{figure}
\begin{center}
\leavevmode
\epsfxsize=14cm
\epsfbox{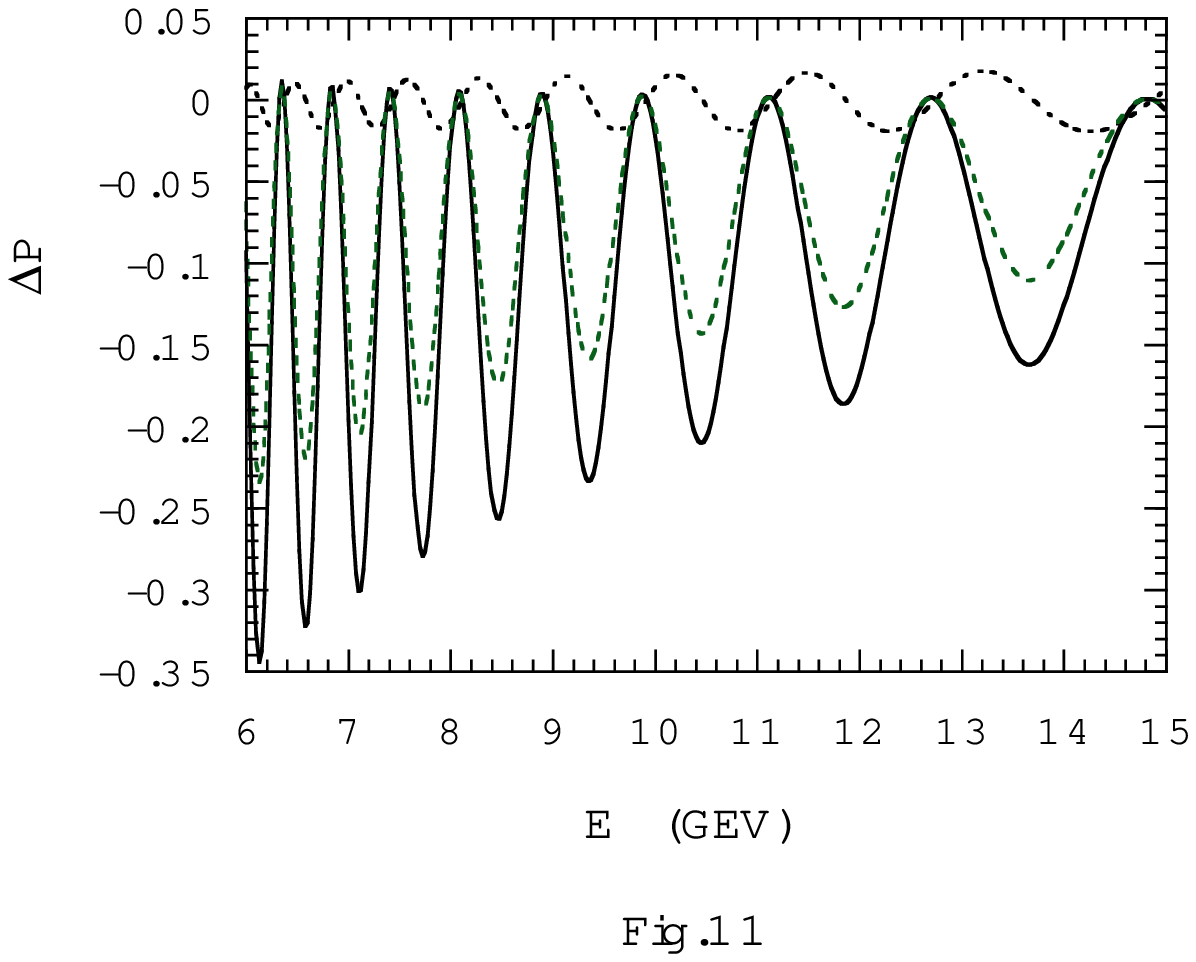}
\end{center}
\caption{The pure CP violation effect in $\nu_{\mu}\to\nu_\tau$
oscillation at $L=730$ km for the typical two parameter sets, $(s_{02}=0.12,
s_{03}=0.06, s_{12}=0.93, s_{13}=0.71, s_{23}=0.75)$(solid line) and
$(s_{02}
=0.12, s_{03}=0.06, s_{12}=0.97, s_{13}=0.73, s_{23}=0.70)$(dashed line) for
the maximal active-sterile admixture $D \sim 0.5$, and the matter effect
(dotted line) for the first parameter set of the above, and commonly taken as
$s_{01}=1/\sqrt{2}, \delta_{01}=\delta_{02}=\delta_{03}=\delta_{12}=0$, and
$\delta_1=\delta_2=\pi /2$.}
\end{figure}
\vskip 0.5truecm

\newpage
\begin{figure}
\begin{center}
\leavevmode
\epsfxsize=14cm
\epsfbox{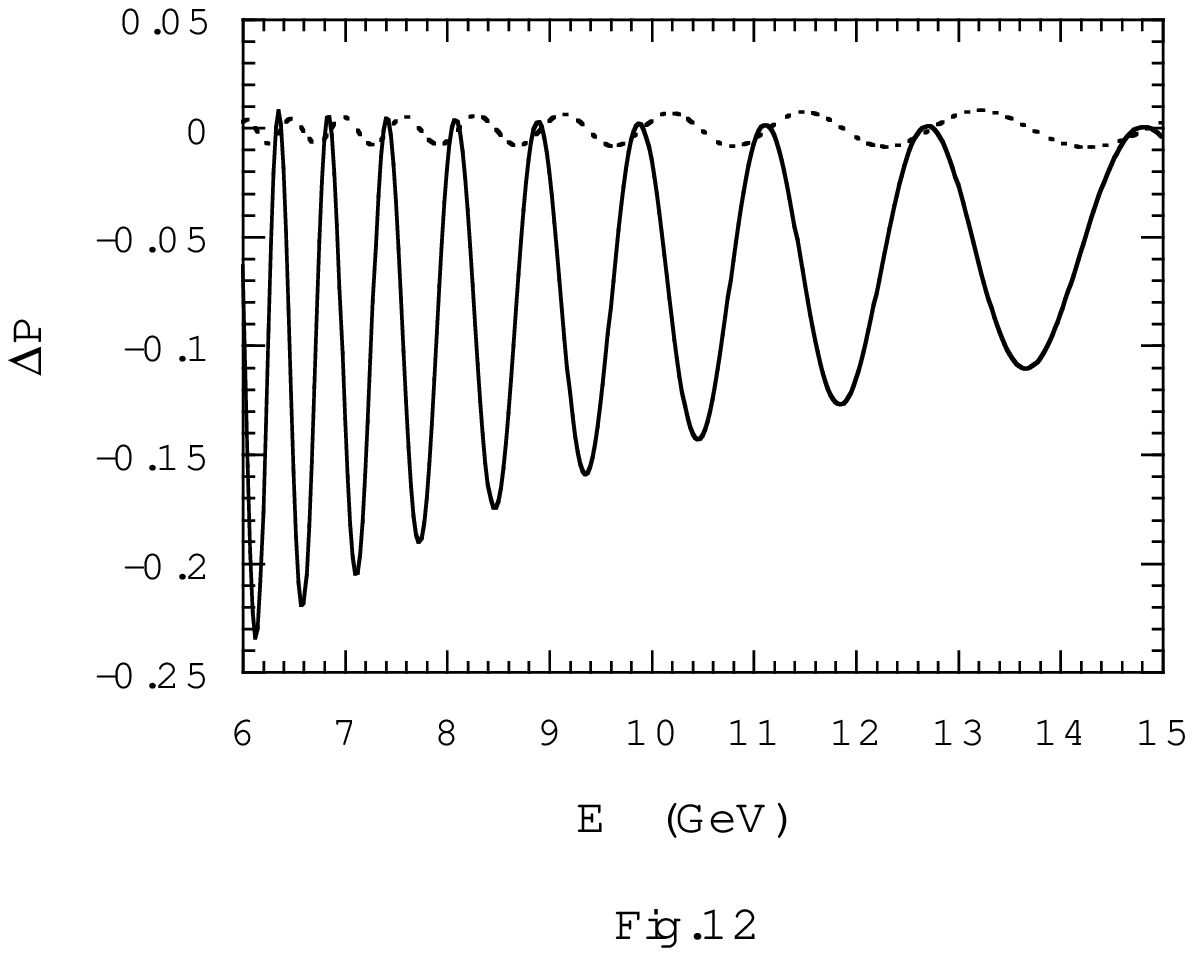}
\end{center}
\caption{The pure CP violation effect (solid line) and the matter effect
(dotted line) in $\nu_{\mu}\to\nu_\tau$ oscillation at $L=730$ km for the
parameter set, $(s_{02}=0.12, s_{03}=0.06, s_{12}=0.97, s_{13}=0.73,
s_{23}=0.70)$ and the other angles and phases are the same as in Fig.11.}
\end{figure}
\vskip 0.5truecm

\end{document}